\documentstyle[seceq,epsf]{ptptex}

\makeatletter
\def\lesim{\mathrel{\mathpalette\gl@align<}}
\def\gtsim{\mathrel{\mathpalette\gl@align>}}
\def\gl@align#1#2{\lower.7ex\vbox{\baselineskip\z@skip\lineskip.2ex%
  \ialign{$\m@th#1\hfil##\hfil$\crcr#2\crcr\sim\crcr}}}
\makeatother

\notypesetlogo  
\title{
 String Theory and the Space-Time Uncertainty Principle}

\author{
Tamiaki {\sc Yoneya}\footnote{  E-mail address: tam@hep1.c.u-tokyo.ac.jp}
}

\inst{
Institute of Physics,  University of Tokyo 
\\
 Komaba, Tokyo 153-8902, Japan
}

\abst{
The notion of  a 
space-time uncertainty principle in string theory 
is clarified and further developed. 
The motivation and the derivation of the principle are first reviewed in a reasonably self-contained 
way.  It is then  shown that the nonperturbative (Borel summed) 
high-energy and high-momentum transfer 
behavior of string scattering is consistent with 
the space-time uncertainty principle. It is also shown  
that, as a consequence of this principle,  
string theories in 10 dimensions generically 
exhibit a characteristic length scale which is equal to 
the well-known 11 dimensional Planck length $g_s^{1/3}\ell_s$ of M-theory  as the scale at which 
stringy effects take over the effects of classical supergravity, even without involving D-branes directly.  The implications of 
 the space-time uncertainty relation in connection with  D-branes and black holes are discussed and reinterpreted.  Finally, we present  
a novel interpretation of the Schild-gauge action 
for strings from the viewpoint of noncommutative 
geometry. This conforms to the space-time 
uncertainty relation by manifestly exhibiting a noncommutativity 
of quantized string coordinates between, dominantly,  
space and time.  We also discuss the consistency  
of the space-time uncertainty relation
with S and T dualities. 
}

\begin{document}
\newcommand{\EQ}{\begin{equation}}
\newcommand{\EN}{\end{equation}}
\newcommand{\EQAN}{\begin{eqnarray*}}
\newcommand{\EQNN}{\end{eqnarray*}}
\newcommand{\EQA}{\begin{eqnarray}}
\newcommand{\EQN}{\end{eqnarray}}
\newcommand{\e}{{\rm e}}
\newcommand{\Sp}{{\rm Sp}}
\newcommand{\Tr}{{\rm Tr}}
\newcommand{\p}{\partial}
\newcommand{\EQAnn}{\begin{eqnarray*}}
\newcommand{\EQNnn}{\end{eqnarray*}}
\newcommand{\boldk}{\mbox{\boldmath $k$}}
\newcommand{\boldp}{\mbox{\boldmath $p$}}
\newcommand{\boldq}{\mbox{\boldmath $q$}}
\newcommand{\boldx}{\mbox{\boldmath $x$}}
\newcommand{\boldy}{\mbox{\boldmath $y$}}
\newcommand{\boldz}{\mbox{\boldmath $z$}}

\maketitle

\section{Introduction}
Since the time that string theory\cite{over} was 
first recognized to be 
the prime candidate for the unified theory 
including gravity \cite{yo0} \cite{ss}, we have been discovering 
a multitude of facets of the theory which increasingly 
reveal its richness  in unexpected ways.  We are more or less convinced that the 
theory must have some hidden but firm foundation behind 
many surprising 
phenomena we observe on its surface. 
However, present string theory  remains essentially 
as  a collection of rules for building the S-matrix 
in perturbation theory.  We do not know 
why such a perturbation theory can arise and 
what the basic  principles leading to the symmetry of string 
perturbation theory are. 
Uncovering the underlying principles of string theory is 
an important necessary step toward a non-perturbative and completely 
well-defined formulation of the theory, based on which we should 
be able to pose various physically relevant questions 
that the ultimate unified theory has to answer,  
but hitherto have not been meaningfully dealt with. 

There have been several  attempts towards 
constructing nonperturbative 
formulations of string theory. The first and most traditional 
one is the field theory of strings, which has been pursued 
quite actively since more than fifteen years ago.\cite{kk}$^{\sim}$\cite{nw}  A related approach 
involves various studies of the geometry of loop space 
and conformal field theories. A notable example is 
an approach based on an abstract complex geometry \cite{fs} 
or, more 
physically, on the renormalization 
group in the theory space of two-dimensional field theories \cite{banksetal}.   
All of these approaches are closely connected  
to each other in a variety of 
ways, depending on the manner in which 
we compare them.   
Curiously enough, however,   it is generally not easy, despite their apparent similarities,  
to establish concrete connections among these attempts at  different formulations 
into a unified scheme in which we can formally go back and 
forth among them.  It seems that, from the viewpoint of representing 
 string amplitudes, the difference between these approaches 
essentially lies in the manners that the moduli space of 
Riemann surfaces is decomposed.\cite{nakanishi}$^{\sim}$\cite{giddmar}  
For example, the gauge invariance 
of string field theory expresses the requirement of 
smooth joining of the decomposed pieces of moduli space. 
The transition among different 
approaches therefore amounts to 
connecting theories with different schemes of the 
decomposition into one single theoretical framework. 
Such a `transformation theory', if 
successfully established, could suggest some crucial 
structure behind string theory by extracting possible 
principles behind conformal symmetry. 

A slightly different approach 
involves the `old' matrix models \cite{oldmatrix} as  toy 
models for studying the theory in which  the entire 
string perturbation series are summed over in lower space-time 
dimensions. Moreover, in recent years,  after the 
discovery of D-branes \cite{pol} and their
 effective description  
in terms of Yang-Mills theory,\cite{wittenym} a new approach that we may 
call the `new' matrix models,  
which can be formulated in higher space-time dimensions, has 
been advocated. Except for some special circumstances,  
such as the infinite-momentum limit and some 
sort of large $N$ scaling limit analogous  to the old matrix models, the new matrix models 
are regarded at best as  effective low-energy 
descriptions of D-branes in terms of the 
lowest string excitation modes alone.  
The new matrix models 
have suggested some unexpected relations 
among local field theories embedded in 
string theory as low-energy approximations.  
The notable exceptions to this interpretation 
of the new matrix models 
might be  the so-called Matrix theory \cite{bfss} and 
the IIB matrix model.\cite{ikkt}  
They have been conjectured to 
be exact theories. However, at the present stage of 
development, it seems fair to say that we have not yet   
succeeded in showing convincingly their validity  as the fundamental formulations of string theory. 

Independently of these specific attempts, one thing is evident. 
Namely, the structure of string theory is governed 
by the conformal symmetry of world-sheet dynamics, 
exhibited by the present perturbative rules.  
Indeed, almost all merits of perturbative 
string dynamics, such as the emergence of the 
graviton, elimination of 
ultraviolet divergences, critical dimensions, 
complete bootstrap between states and interactions, and so on,   
are direct consequences of the world-sheet conformal 
invariance.  This is true even when we take into account 
various brane excitations, since the 
interactions of the branes are mediated by the 
exchange and fluctuations 
 of  strings.  
The motion and 
interaction of D-branes are described by the 
ordinary elementary (or `fundamental') strings and their 
vertex operators in terms of the world-sheet dynamics. 

The exact conformal invariance of  world-sheet 
dynamics of strings actually corresponds to 
the fact that the genuine observables of string 
theory are solely on-shell S-matrix elements. 
Given only  an S-matrix, however, it is in general not 
easy to determine the real symmetries and 
degrees of freedom which are appropriate to 
express the content of the theory off shell.  
In particular, it is not obvious what is the 
appropriate generalization of the 
world-sheet conformal invariance to the 
off-shell formulation of string theory. 
This problem is a long-standing example among the 
several major 
obstacles we encounter in trying 
to formulate the nonperturbative dynamics 
of string theory. 
The nature of the problem is somewhat reminiscent 
of that which characterized the well-known history of physics evolving from early quantum theory to quantum mechanics.  
The quantized atomic spectra were 
derived by imposing the Bohr-Sommerfeld quantization 
conditions, characterized by the adiabatic 
invariance, which select particular orbits  
from the continuous family of allowed orbits in the 
phase space of classical mechanics.  
Then quantum mechanics replaced the 
Bohr-Sommerfeld condition by a deeper and universal structure, 
namely, Hilbert space and the algebra of observables defined on it, characterized by 
the superposition principle and the canonical commutation relations, respectively.  
We should perhaps expect something similar 
in string theory. Namely, 
the condition of conformal invariance, as the analog 
of the adiabatic invariance of the quantum condition,  
 must be generalized to 
a more fundamental and universal structure which allows us 
to construct the concrete nonperturbative theory.  

However, 
in view of the status of past 
approaches to nonperturbative definitions  
of string theory as briefly summarized above, 
it seems that the correct language and 
mathematical framework 
for formulating string theory remain yet to be discovered. 
Given this situation, it is worthwhile to 
attempt extracting the most characteristic qualitative 
properties of string theory originating from the 
world-sheet conformal invariance 
which are universally valid, irrespective  
of the particular formulations of string theory. 
In particular, in order to seek some clear 
basis behind conformal symmetry, 
it seems more advantageous to 
express directly the smooth nature of the 
moduli space of the Riemann surface without 
making decompositions of it, as is usually done in string field 
theories, for example.

Although such universal qualitative properties 
may, by definition, be too crude to make   
any quantitative predictions,  as opposed to simply giving 
rough order estimates for some simple and typical 
phenomena, 
they might be of some help in pursuing 
the underlying principles of string theory 
{\it if} they characterize critically the departure  of string theory 
 from the physics governed 
by the traditional framework of local quantum field theories.  
The proposal of a space-time uncertainty principle \cite{yo}$^,$\cite{yo1} was motivated by this manner of thinking.  
The purpose of the present paper is to clarify and further develop 
the space-time uncertainty principle of string theory 
from a new perspective.  We first review the original derivation 
with some clarifications in \S 2.  We also make 
comparison of our space-time uncertainty relation 
with other proposals of similar nature, such as the 
notion of a modified uncertainty relation with stringy 
corrections.  We explain why 
the latter cannot be regarded as a universal relation 
in string theory,  
in contrast to our space-time 
uncertainty relation,   which we
argue here to be valid nonperturbatively. 
Its implications are 
then discussed with regard to some aspects of 
string theory, mainly the high-energy limit 
of the string S-matrix in \S 3,  and  the characteristic physical 
scales in the physics of microscopic black holes 
and D-particle scatterings in \S 4.  It is argued that 
the high-energy behavior with fixed scattering angle  almost saturates  the space-time uncertainty relation 
after summation 
over the genus by the Borel sum technique. 
This suggests strongly that the 
space-time uncertainty relation is valid nonperturbatively, being obeyed 
independently of the strength of the string coupling, at 
least for a certain finite range of the 
string coupling, including the 
weak coupling regime.  
It is shown that the M-theory scale is a natural consequence 
of the space-time uncertainty relation combined with 
the properties of microscopic black holes, even without invoking D-branes directly.  
The saturation of the space-time uncertainty 
relation is also shown to be one of the characteristic 
features of D-particle scatterings. It is argued that 
D-particle and anti-D-particle scattering, in general, 
do not saturate the space-time uncertainty relation. 
\S 4 also contains some remarks on the possible roles 
of the space-time uncertainty relation in connection 
with some developments 
of string theory, such as black-hole complementarity,  
holography and UV-IR correspondence. In particular, 
a simple interpretation of the Beckenstein-Hawking entropy 
for the macroscopic Schwarzschild black hole 
is given from the viewpoint of the space-time 
uncertainty relation. 
Then, in \S 5 we proceed to suggest the  possibility 
of formulating the space-time uncertainty principle 
by quantizing string theory 
in a way which conforms to noncommutative geometry, 
exhibiting  manifestly a noncommutativity between space and time.  
The argument is based on a 
novel interpretation of the string action in the Schild gauge, 
but its completion toward a concrete calculable 
scheme is left for future works.  
The final section is devoted to discussion of some 
issues which are not treated in the main text,  
including the interpretation of S and T 
dualities from the viewpoint of the space-time 
uncertainty principle, 
and of 
future prospects. 

In addition to developing the ideas of the 
space-time uncertainty relation further, it is another purpose 
of the present paper to discuss most of the various relevant issues  
in a more coherent fashion from a definite standpoint, 
since previous discussions regarding the 
space-time uncertainty principle, mainly in 
works by 
the present author and including 
some works by other authors,  are 
scattered in various different papers.
We would like to lay a foundation for further 
investigation by discussing their meaning and usefulness 
in understanding the nature of string theory.  
The present author also hopes that this exposition 
will be useful to straighten up some 
confusion and awkward prejudices prevailing in the literature 
and to clarify the standpoint and, simultaneously, 
the limitations and remaining problems 
of the present qualitative approach. By so doing, 
we may hope to envisage some hints of a 
 truly satisfactory 
formulations of the basic principles of string theory.

\section{Derivation of the space-time uncertainty 
relation}
 
The first proposal \cite{yo} of the space-time uncertainty relation 
in string theory came from an elementary space-time  interpretation of 
the mechanism of eliminating ultraviolet divergencies in string theory. 
As is well known, the main reason that string 
amplitudes are free from ultraviolet divergences 
is that the string loop amplitudes satisfy the 
so-called modular invariance.  The latter symmetry, 
which is a remnant of the conformal 
symmetry of Riemann surfaces after a gauge 
fixing, automatically 
generates the cutoff for the short-distance parts  
of the integrations over the proper times 
of the string propagation in loop diagrams.  
In traditional field theory approaches, 
the introduction of an ultraviolet cutoff 
 suffers, almost invariably, 
from the violation of unitarity and/or locality. 
However, string perturbation theory 
is perfectly consistent with (perturbative) unitarity, 
preserving all the important axioms for a physically 
acceptable S-matrix, including the analyticity property of the S-matrix. 
It should be recalled that the analyticity of  the 
S-matrix is customarily attributed to locality, in addition to 
unitarity,  of quantum 
field theories.  However, locality is usually not expected to be 
valid  in theories with extended objects. 
From this point of view, 
it is not at all trivial to understand 
why string theory is free from the 
ultraviolet difficulty,  and it is important to give 
 correct interpretations to its mechanism.  

\subsection{A reinterpretation of energy-time 
uncertainty relation in terms of strings}

The approach which was  proposed in Ref.\cite{yo} 
is to reinterpret the ordinary 
energy-time  uncertainty relation in terms of the 
space-time extension of strings:\footnote{
Throughout the present paper, we use units in which $\hbar =1, c=1$. 
} 
\EQ
\Delta E \Delta t \gtsim 1 .
\label{energytime}
\EN
The basic reason why we have ultraviolet 
divergencies in local quantum field theories is that 
in the short time region, $\Delta t \rightarrow 0$, 
the uncertainty with respect to energy increases 
indefinitely: $\Delta E 
\sim 1/\Delta t\rightarrow \infty$. This 
in turn induces a large uncertainty in momentum 
$\Delta p \sim \Delta E $. The large uncertainty 
in the momentum implies
 that the number of particles states allowed in the 
short distance region $\Delta x \sim 1/\Delta p$ 
grows indefinitely as $(\Delta E)^{D-1}$  in 
$D$-dimensional space-time. In ordinary 
local field theories, where 
there is no cutoff built-in, all these states 
are expected to 
contribute to amplitudes with equal strengths. 
This consequently leads to UV infinities.  

What is the difference, in string theory, 
regarding this general argument?  
Actually, the number of the allowed states 
with a large energy uncertainty $\Delta E$ behaves as 
$e^{k \ell_s \Delta E}\sim e^{k \ell_s/\Delta t}$ with some positive coefficient $k$,  
and  $\ell_s \propto \sqrt{\alpha'}$ being the string length constant,  
where $\alpha'$ is the traditional slope parameter. 
This increase of the degeneracy  is much faster  
than that in local field theories.  The crucial difference with local field theories, however,  is 
that the dominant string states among these exponentially 
degenerate states are not the states with large 
center-of-mass momenta, but can be the massive states 
with higher excitation modes along  strings. 
The excitation of higher modes 
along strings contributes to the large spatial extension 
of string states. It seems reasonable to expect that this effect completely cancels the  
short distance effect with respect to the 
center-of-mass coordinates of strings, provided that 
these higher modes contribute  appreciably to physical 
processes.  
Since the order of  magnitude of 
the spatial extension corresponding to a 
large energy uncertainty $\Delta E$ is expected to behave as 
$\Delta X \sim  \ell_s^2 \Delta E$, we are led 
to a remarkably simple 
relation for the order of magnitude $\Delta X$ 
for fluctuations along spatial directions of string states participating within the 
time interval $\Delta T=\Delta t$ of interactions: 
\EQ
\Delta X \Delta T \gtsim \ell_s^2 .
\label{stu}
\EN
It is natural to call this relation the `space-time 
uncertainty relation'.  It should be emphasized that this 
relation is {\it not} a modification of the usual 
uncertainty relation, but simply a {\it reinterpretation} 
in terms of strings. Note that as long as 
we remain in the framework of quantum mechanics, 
the usual Heisenberg relation $\delta x \delta p \gtsim 1$ is 
also valid if it is correctly interpreted. For example, 
the latter is always valid for the center-of-mass momentum 
and the center-of-mass position of strings. The space-time uncertainty 
relation, on the other hand, gives a new restriction 
on the short-distance space-time structure, which 
comes into play because of the intrinsic extendedness 
of strings. In general, therefore, we have 
to combine the ordinary uncertainty relation and 
the space-time uncertainty relation appropriately in 
estimating the  relevant scales in string theory, 
as is elucidated in  discussion given below.  

To avoid a possible misunderstanding, we remark that, 
as is evident in this simple derivation, the 
spatial direction is dominantly along the longitudinal direction 
of strings. Therefore, it should not be 
confused as the more familiar transverse spread 
of a string. 
That the longitudinal size indeed grows linearly with energy,  
at least in perturbation theory 
is most straightforwardly seen as follows. 
For simplicity, let us consider the case of open strings. 
The interaction of open strings is 
represented by the vertex operator $\exp ip x(\tau, 0)$, 
inserted as the endpoint $\sigma =0$ of one of the strings. 
If the string before the insertion is made 
is in the ground state with some moderate momentum, 
the effect of the vertex operator is to change the 
state after the insertion to a coherent state of the 
form $\exp (p\alpha_{-n}\ell_s/n)|0\rangle$ for each 
string mode $n$.  This induces a contribution to the spatial extension of the  string coordinate along the 
spatial components $\vec{p}$ of the momentum vector 
$p_{\mu}$ of order $\langle x_n^2 \rangle \sim 
|\vec{p}|^2 \ell_s^4 /n^2$, which in the high-energy limit $
|\vec{p}| \rightarrow \infty $ leads 
to $\Delta X \sim \sqrt{|\vec{p}|^2 \ell_s^4\sum_{n<n_s} (1/n^2)}\sim E\ell_s^2$.  Note that 
here we have adopted as the measure of string extension 
the quantity $\sqrt{\langle\int d\sigma \, (x(\sigma)-x_0)^2\rangle}$. 
This apparently shows that 
there is a large extension with respect to the time direction too. 
However, the interaction time $\Delta T$ should be defined with respect to 
the center-of-mass coordinate of strings, and hence 
the apparent 
large extension along the time direction does not 
directly correspond to the time uncertainty in the 
energy-time uncertainty relation (\ref{energytime}). 
Furthermore, 
we should expect the existence of some limit 
$n \lesim n_{s}$ 
for the  excitation of string modes, depending on 
the specific region that the string scattering is probing. 
In the Regge limit, for example, 
 where the momentum transfer is 
small, we can show that $\ell_s/\Delta T\sim n_s \lesim s\ell_s^2$ 
(see \S 3.3). In this case, 
in addition to the growth along the longitudinal direction, 
we have an intrinsic transverse extension of 
order $\ell_s\sqrt{\sum_{n \lesim n_s}(1/n)}\sim \ell_s\sqrt{\log (E\ell_s)}$ for all directions 
corresponding to the zero-point extension of the ground state 
wave function.  The logarithmic transverse extension 
is negligible compared to the linear growth. 
The mechanism of suppressing the 
ultraviolet divergence, as exhibited by the modular 
invariance, cannot be attributed to the logarithmic growth of 
the extension of the ground state wave function. 
It is clearly the effect that is dominantly 
associated with the longitudinal extension of strings.  

We will later see that in some cases, such as the case of high-energy-high-momentum transfer scattering of strings 
and D-particle scattering with small velocities, 
we can effectively 
neglect the effect of  string higher modes. 
This is not directly in contradiction with the role of 
string higher modes which we 
emphasized above in connection with the 
enormous degeneracy of string states associated with the 
higher modes.  
The degeneracy is with regard to the standard string modes 
of free strings with standard boundary conditions. 
Situations in which the string higher modes are 
effectively negligible occur with 
different backgrounds or different boundary conditions 
for the string coordinates as fields on the string world sheet. 
In terms of the standard free strings, such 
cases are represented by a coherent state with 
excitation of higher string modes. 

The main purpose of the present paper is to present 
 several arguments which suggest that 
the space-time uncertainty relation 
(\ref{stu})  may be a universal principle 
which is valid nonperturbatively in string theory.  
It should be emphasized that  
the space-time uncertainty principle has yet 
only been qualitatively formulated.  
We cannot give a rigorous definition for the 
uncertainties $\Delta X$ and $\Delta T$ at the 
present stage of development.  
For example, one might ask how to define the 
time uncertainty if the string stretches linearly with 
energy. We always assume that the time is 
measured with respect to some preferred point, most 
naturally at the center-of-mass of a string. 
Also, there is no point-like probe with which we can 
measure the spatial uncertainty of a string: 
A string itself has an intrinsic extension depending 
on the scale of resolution if we are allowed to
 imagine a point-like probe. 
  
The point we would like to stress is, however, that 
this simple looking relation has quite universal applicability 
both perturbatively and nonperturbatively,  at least 
in some qualitative sense, 
if it is interpreted appropriately.  
Also, involving both time and space intrinsically, 
the relation is not just a kinematical 
constraint which decreases the number of degrees of 
freedom, but in principle may place a strong 
constraint on the dynamics of the system. 
Its precise role and the correct mathematical 
formulation can only be  
 found after the proper  
framework of string theory is established.  
The prime motivation for this viewpoint was a general belief 
that any theory of quantum gravity 
must impose some crucial restrictions at short distance scales near the Planck length,  beyond which  
the classical space-time geometry,  on 
which general relativity is based on, is invalidated.  
It is then important to ascertain how 
such a restriction is realized in string theory, 
since it would suggest the precise manner in which  
 string theory  departs from the 
usual framework of quantum field theories.  
The present author is aware of  many past 
attempts at the construction of a formal `space-time quantization'. 
The standpoint in proposing the 
space-time uncertainty principle 
is not to propose yet another version 
of the formal theory of quantized space-time, but to 
possibly uncover some secrets 
that would help lead to the unification of quantum theory and 
general relativity  from string theory which possesses  several  ideal properties as a candidate for the unified theory in 
a quite surprising and unexpected fashion.  
 
\subsection{The nature of the space-time uncertainty relation} 

Now an important characteristic of the relation (\ref{stu}) 
is that it demands a duality between the time-like and 
space-like distance scales.  Whenever we attempt to probe the 
short distance region $\Delta T\rightarrow 0 $ in a time-like direction, 
the  uncertainty with respect to the space-like direction 
increases. In addition,  we propose that the 
relation is also valid in the opposite limit. 
Namely, if we attempt to probe the short distance region 
$\Delta X\rightarrow 0$ in space-like directions, then the uncertainty 
$\Delta T$ in the time-like direction increases. 
In other words, 
the smallest distances  probed in string theory cannot be 
made arbitrarily small with respect to 
both time-like and space-like directions 
{\it simultaneously}.  
It is proposed in \cite{yo1} that 
the phenomena of minimal distance \cite{minimal}
\cite{amati}  in string perturbation theory can be interpreted in this way.  However, it should be kept 
in mind that our space-time uncertainty relation does not forbid the 
possibility of probing shorter distance regions than the 
ordinary string scale in string theory, 
quite contrary to the implication following
 the usual notion of 
minimal possible distance in string theory.\footnote{
The possible relevance of shorter length scales has been 
later suggested in Ref.\cite{shenker}. 
} 
Rather, it only imposes a new condition 
that the short and large 
distances are dual to each other.  
We note that in some of the recent developments of 
nonperturbative string theory associated with 
D-branes, the regime of short open strings much below the 
string scale is a crucial ingredient.  

What is the physical interpretation of the opposite limit, namely the short spatial distance which implies a 
large time uncertainty $\Delta T \rightarrow 
\infty$? Is it really possible to probe 
distance scales $\Delta X$ smaller than $\ell_s$? 
Since any string state with a definite mass has 
an intrinsic spatial extension of order $\ell_s$, it 
seems at first sight impossible to do this. It turns out that 
the D-particle, instead of the 
fundamental strings, plays  precisely 
such a role as shown later. 
Moreover, the fact that the asymptotic 
string states can be represented by vertex operators 
coupled with local external fields may be interpreted as 
a consequence of the relation $\Delta X \sim \ell_s^2/\Delta T \rightarrow 0$.  In this sense, the space-time 
uncertainty relation can also be viewed as a 
natural expression of the $s$-$t$ duality, which has 
been the basic background for string theory. 
Roughly speaking, the resonance poles near on-shell in the $s$-channel correspond to 
$\Delta T\rightarrow \infty$, while 
the $t$-channel massless pole exchange to $\Delta X\rightarrow 
\infty$, with vanishingly small 
momentum transfer,  
if the  exchange is interpreted in terms of pair 
creation and annihilation of open strings. 
In fact, the $s$-$t$ duality was another motivation for 
proposing the space-time uncertainty relation 
in Ref. \cite{yo1}.

The fact that the string theory has  a short distance cutoff 
builtin in this way 
might be somewhat counter intuitive, since strings have  
a much larger number of particle degrees of 
freedom  than any local field 
theories or ordinary nonlocal field theories 
with  multi-local fields and/or some nonlocal interactions. 
But precisely because of this counter-intuitive nature 
of string theory, we must study the short distance structure 
of string theory carefully.  
For example, the growth of the string size along the 
longitudinal direction with energy might seem to be quite 
contrary with the familiar idea of Lorentz 
contraction of a projectile. However, this is 
one of the properties that allows, 
at least in perturbation theory, 
string theory to contain gravity, 
as we will discuss in \S 3. 
Also, the large degeneracy of particle states  should rather be interpreted as implying that string theory 
suggests an entirely new way for counting the physical 
degrees of freedom in the region of
 the smallest possible distance scales. 
We hope that the discussion given here will provide 
a basis for the concrete formulation of 
this general idea. 
 
Before proceeding further, it is appropriate here to 
comment on the difference between our space-time 
uncertainty relation and the 
other proposal of a related uncertainty relation with stringy corrections. 
In parallel to the original suggestion\footnote{
Unfortunately, since the proposal (\ref{stu}) was initially  made in a 
paper \cite{yo} written for the volume commemorating  
Prof. Nishijima's 60th birthday and has not been 
published in 
popular journals, it has long 
been ignored.   The earliest 
discussion of the space-time uncertain relation 
in the popular journals was presented in Ref.\cite{yo1}. It, however, seems that  even this reference 
has been largely ignored to date.   The author hopes that 
the present exposition is useful in drawing attention to these 
papers. } of the 
space-time uncertainty relation,   the 
high-energy behavior of the string amplitudes 
has been studied.  On the basis of such investigations, it  was proposed independently of the proposal (\ref{stu}) that 
in the high-energy limit  
the space-time  extensions of strings are proportional \cite{grossmende} 
to energy and momentum as\[
x^{\mu} \propto \ell_s ^2\, p^{\mu} .
\]
The reason behind this proposal is that the classical 
solution for the string world-sheet 
trajectory for  given wave functions with momenta $p_i^{\mu}$ corresponding 
to external asymptotic states  takes the following 
form  
\EQ
x^{\mu}(z, \overline{z})=
\ell_s^2 \sum_i p^{\mu}_i \log |z-z_i| 
\label{classicalworldsheet}
\EN
in the lowest tree approximation, where the $z_i$'s are the positions of the vertex operators 
on the Riemann surface 
corresponding to the asymptotic states with on-shell 
momenta $p^{\mu}_i$. 
This seems also to be consistent with what we have 
discussed using the vertex operator in our 
derivation of the space-time uncertainty relation. 
Combined with the Heisenberg 
relation $|\delta x^{\mu}| \sim 1/|\delta p^{\mu}|$, 
the above proposal 
suggests a modified uncertainty relation \cite{moduncer} for each 
space-time component (no summation over $\mu$), 
\EQ
|\delta x^{\mu}||\delta p^{\mu}|  \gtsim 1 + \ell_s^2 \, |\delta p^{\mu}|^2 ,
\label{stringyuncertainty}
\EN
which leads to $|\delta x^{\mu}| \gtsim \ell_s$ for 
{\it each} component of the space-time coordinates 
separately.   Our space-time 
uncertaity relation (\ref{stu}) is weaker than 
this relation, and it does not  lead directly to 
the minimul distance, unless we assume some 
relation between time and space uncertainties : 
For example, if we set $\Delta T \sim \Delta X$, we 
immediately have the miminum distance relation 
$\Delta X \gtsim \ell_s$.  This is a crucial difference. 

It appears that this particular form (\ref{stringyuncertainty}) cannot be regarded as being 
universally valid in string theory. 
We can provide at least three reasons for this. 
First, the uniform proportionality between energy-momenta and the extensions 
of the string coordinates  is not 
valid in the region in which the
high-energy behavior is dominated by the 
Riemann surfaces where the positions of the vertex operators 
approach the boundary of the moduli space.  
 Second, even when the dominant contribution comes from a 
region which is not close to the boundary of the 
Riemann surface, it is known that the 
amplitudes after summing up the entire perturbation series 
using the Borel-sum technique  
behave differently from the tree approximation for high-energy fixed angle scattering. 
The known behavior is incompatible with the relations such as (\ref{stringyuncertainty}) demanding that the 
string extension grows indefinitely, while it turns out to be consistent with 
our relation (\ref{stu}).  
Third and most importantly, the relation (\ref{stringyuncertainty}) 
is not effective for explaining the short-distance 
behavior of D-brane interactions. In particular, 
a naive relation such as $|\delta x^{\mu}| \gtsim \ell_s$,   expressing  the presence of a minimal distance,  
  clearly contradicts the decisive role of the familiar 
characteristic spatial 
scale $g_s^{1/3}\ell_s$  
in D-particle scattering, which is much 
smaller than the string length $\ell_s$ 
in the weak-coupling regime, and more generally in 
the conjecture of M-theory \cite{wittenm}. 
All of these points will be discussed fully in later sections. 

One might naively think that when
 the spatial extension becomes 
large the interaction time would also increase, as expressed 
in (\ref{stringyuncertainty}), 
since the spatial region for interaction grows. 
This intuition might be correct if we were dealing with 
ordinary extended objects, such as polymers, which 
may interact with each other in the bulk of the spatial 
extension. However, the nature of the interaction of 
elementary strings is strongly constrained by 
conformal invariance.  Elementary strings have no bulk-type forces among their parts. Thus, the ordinary 
intuition for the extended object is not necessarily 
applicable to string theory.  For this reason, whether 
the interaction time should also increase as 
the spatial extension or not must depend on 
specific situations and cannot be stated as a general 
property. 

As the final topic of this subsection, let us ask   
whether and how the space-time uncertainty relation (\ref{stu}) 
can be compatible with kinematical Lorentz invariance.  
 The answer is that the relation as an {\it inequality} can be 
consistent with Lorentz invariance.  Suppose that the 
relation is satisfied in some preferred Lorentz frame which 
we call the proper frame, where the uncertainties are 
$\Delta T= \Delta T_0$ and $\Delta X = \Delta X_0$, and, 
in particular, the spatial uncertainty can be estimated 
as being at rest.  In most physical 
applications discussed later in this paper, we always 
assume such a preferred frame in deriving the 
relation.  Let us make a Lorentz boost of the frame of 
reference with velocity 
$v$ and measure the same lengths in the boosted frame. 
Then the uncertainty in time is 
$\Delta T=\Delta T_0 /\sqrt{1-v^2}$, while 
the spatial interval is contracted as 
$\Delta X =\sqrt{1-v^2}
\Delta X_0$ or is not affected $\Delta X=\Delta X_0$ depending 
on the directions of the characteristic spatial scale. 
This shows that the inequality (\ref{stu}) is 
preserved in any Lorentz frame provided that 
it is satisfied in some proper Lorentz frame, after averaging 
over the spatial directions.

We can arrive at the same conclusion  from more formal 
considerations too. Let us  temporarily suppose the existence of 
an algebraic  realization of the space-time uncertainty relation, 
by introducing the space-time (hermitian) operators $X^{\mu}$, transforming as 
Lorentz vectors,  as some effective agents 
measuring the observable distance scales in 
each Minkowski direction $\mu$.  Then,  as 
has been discussed in a previous paper,\cite{yo3} 
 the space-time 
uncertainty relation may correspond to an 
operator constraint which is manifestly Lorentz 
invariant, as given by
\EQ
{1\over 2}[X^{\mu}, X^{\nu}]^2\sim \ell_s^4 ,
\label{cconstraint}
\EN
where the contracted indices are summed over as usual.\footnote{
Similar constraints are studied from a different 
viewpoint in Ref.\cite{dopl}. 
} 
By decomposing into time and space components, 
we have 
\EQ
\sqrt{\langle -[X^0, X^i]^2 \rangle} = \sqrt{{1\over 2} \langle 
-[ X^i, X^j]^2 \rangle + \ell_s^4} \gtsim \ell_s^2 .
\EN
 This shows that the inequality (\ref{stu}) of the 
space-time uncertainty relation can in principle be 
consistent with Lorentz invariance, conforming to 
the first argument.  This also suggests a possible 
 definition of the  proper frame such that the 
noncommutativity of space-like operators 
is minimized.  To avoid a possible misconception, however, 
it should be noted that the present formal argument 
is {\it not} meant to imply 
that the author is proposing that the 
operator constraint (\ref{cconstraint}) 
is the right way for realizing the 
space-time uncertainty principle.  
In particular, 
it is not at all obvious whether the 
uncertainties can be defined using Lorentz 
vectors, since they are not local fields. 
Here it is only 
used for an illustrative purpose to  
show schematically the compatibility of the space-time 
uncertainty relation with Lorentz invariance.  
There might be better way of formulating 
the principle in a manifestly 
Lorentz invariant manner.  We will later present a related discussion (\S 5) 
 from the  viewpoint of  noncommutative geometric 
quantization of strings based on the Schild action.

\subsection{Conformal symmetry and the 
space-time uncertainty relation}

For the sake of completeness,  we now explain an independent derivation of  the space-time uncertainty relation on the basis of conformal invariance of the world-sheet string dynamics, following an old work.\cite{yo1} This derivation seems to support our 
proposal that the 
space-time uncertainty relation should be valid universally 
in both short-time and long-time limits. 

All the string amplitudes are formulated in terms of path integrals as weighted mappings from the set of all possible Riemann surfaces to a target space-time. Therefore, any characteristic 
property of the string amplitudes can be 
understood from the property of this path integral. 
The absence of the ultraviolet divergences 
in string theory from this point of view 
is a consequence of the modular invariance. 
We will see that the space-time uncertainty 
relation (\ref{stu}) can be regarded as a natural generalization  of the modular invariance for arbitrary string 
amplitudes in terms of the 
direct space-time language. 

Let us start by briefly recalling how to define the 
distance on a Riemann surface in a conformally 
invariant manner.  For a given Riemannian metric 
$ds = \rho(z,\overline{z}) |dz|$, an arc $\gamma$ on the Riemann surface has  length $L(\gamma, \rho)
=\int_{\gamma}\rho |dz|$. This length is, however,  
dependent on the choice of the metric function $\rho$. 
If we consider some finite region $\Omega$ and 
a set of arcs defined on $\Omega$, the following definition, called the `extremal length' 
in mathematical literature,\cite{alf} is known to give a conformally 
invariant definition for the length of the set $\Gamma$ of arcs: 
\EQ
\lambda_{\Omega}(\Gamma) =
\sup_{\rho}{L(\Gamma, \rho)^2\over A(\Omega, \rho)}
\label{extremallength}
\EN  
with 
\[
L(\Gamma, \rho)=\inf_{\gamma\in \Gamma} L(\gamma, \rho) ,  \quad 
A(\Omega, \rho)=\int_{\Omega}\rho^2 dzd\overline{z} .
 \]
Since any Riemann surface 
corresponding to a string amplitude can be 
decomposed into a set of 
quadrilaterals pasted along the boundaries 
(with some twisting operations, in general), it is 
sufficient to consider the extremal length for an arbitrary  quadrilateral segment $\Omega$. Let the two pairs of 
opposite sides of $\Omega$ be $\alpha, \alpha'$  and 
$\beta, \beta'$. Take $\Gamma$ be the set of all 
connected set of arcs joining $\alpha$ and $\alpha'$. 
We also define the conjugate set of arcs $\Gamma^*$ be 
the set of arcs joining $\beta$ and $\beta'$. 
We then have two extremal distances, 
$\lambda_{\Omega}(\Gamma)$ and $\lambda_{\Omega}
(\Gamma^*)$.  The important property of the extremal 
length for us is the reciprocity
\EQ
\lambda_{\Omega}(\Gamma)\lambda_{\Omega}(\Gamma^*)=1 .
\label{reciprocity}
\EN
Note that this implies that one of the two mutually 
conjugate extremal lengths is larger than 1. 

The extremal lengths satisfy the 
composition law, which partially  justifies the naming 
``extremal length": 
Suppose that $\Omega_1$ and $\Omega_2$ are disjoint 
but adjacent open regions on an arbitrary Riemann surface. Let $\Gamma_1$ and $ \Gamma_2$ consist of arcs in 
$\Omega_1 $ and $ \Omega_2$, respectively.
Let $\Omega$ be the union $\Omega_1+\Omega_2$,  
and let $\Gamma$ be a set of arcs on $\Omega$. 
\begin{enumerate}
\item If every $\gamma \in \Gamma$ contains a $\gamma_1
\in \Gamma_1$ and $\gamma_2\in \Gamma_2$, then 
\[
\lambda_{\Omega}(\Gamma)\ge \lambda_{\Omega_1}(\Gamma_1) 
+ 
\lambda_{\Omega_2}(\Gamma_2) .
\]
\item If every $\gamma_1\in \Gamma_1$ 
and $\gamma_2\in \Gamma_2$ contains a 
$\gamma\in \Gamma$, then 
\[
1/\lambda_{\Omega}(\Gamma)\ge 1/\lambda_{\Omega_1}(\Gamma_1) 
+ 
1/\lambda_{\Omega_2}(\Gamma_2) .
\]
\end{enumerate}
These two cases correspond to two different types 
of compositions of open regions, depending on whether the side 
where $\Omega_1$ and $\Omega_2$ are joined does not divide 
 the sides which $\gamma\in \Gamma$ connects, or do divide,  
respectively.  
One consequence of the composition law is that 
the extremal length from a point to any finite 
region is infinite and the corresponding conjugate length 
is zero. This corresponds to the fact that the vertex operators 
describe the on-shell asymptotic states whose coefficients  
are represented by  local external fields in space-time. 
We also recall that the moduli parameters of world-sheet 
Riemann surfaces are nothing but a set of 
extremal lengths with some associated 
angle variables, associated with twisting 
operations,  which are necessary in order to specify the joining of the boundaries of quadrilaterals.  

Conformal invariance allows us to conformally map 
any quadrilateral to a rectangle on the Gauss plane. 
Let the Euclidean lengths of the sides 
$(\alpha, \alpha')$ and $(\beta, \beta')$ be $a$ and $b$, 
respectively.  Then, the extremal lengths are given  by the ratios 
\EQ
\lambda(\Gamma)=a/b ,\quad \lambda(\Gamma^*)=b/a. 
\EN
For a proof, see Ref.\cite{alf}
 
Let us now consider how the extremal length is reflected by 
 the space-time structure probed by general string amplitudes. 
The euclidean path-integral in the conformal gauge 
is essentially governed by the action 
$
{1\over \ell_s^2} \int_{\Omega} dzd\overline{z} \, 
\partial_z x^{\mu}\partial_{\overline{z}} x^{\mu}. 
$
Take a rectangular region as above and the boundary 
conditions $(z=\xi_1 + i \xi_2)$ as  
\EQA
x^{\mu}(0, \xi_2)=x^{\mu}(a, \xi_2)=\delta^{\mu 2} B \xi_2/b ,
&& \nonumber \\
x^{\mu}(\xi_1, 0)=x^{\mu}(\xi_1, b)=\delta^{\mu 1} A \xi_1/a \nonumber.
\EQN
The boundary conditions are chosen such that the kinematical 
momentum constraint $\partial_1x^{\mu}\cdot 
\partial_2 x^{\mu}=0$€
in the conformal gauge is satisfied for the classical solution.\footnote{
The Hamilton constraint $(\partial_1 x)^2=(\partial_2x)^2$ 
is satisfied after integrating over the moduli parameter, which  
in the present case of a rectangle is the extremal length itself. 
}
The path integral then contains the factor 
\[
\exp\Bigl[
-{1\over \ell_s^2}\Bigl(
{A^2\over \lambda(\Gamma)} + {B^2\over \lambda(\Gamma^*)}
\Bigr)
\Bigr] .
\]
This  indicates that the square root of extremal length 
can be used as the measure of the length 
probed by strings in space-time. The appearance of the 
square root is natural, as suggested from the definition 
(\ref{extremallength}):
\[  
\Delta A =\sqrt{\langle A^2} \rangle \sim \sqrt{\lambda(\Gamma)}\ell_s , \quad 
\Delta B =\sqrt{\langle B^2\rangle }\sim \sqrt{\lambda(\Gamma)}\ell_s. 
\]
In particular, this implies that  probing short distances 
along both directions simultaneously 
is always restricted by the 
reciprocity property (\ref{reciprocity}) of the extremal length, 
$\Delta A \Delta B \sim \ell_s^2$. 
In Minkowski metric, one of the directions is 
time-like and the other is space-like, as required 
by the momentum constraint. 
This conforms to the space-time uncertainty relation,  
as derived in the previous subsection from a very 
naive argument. 
Also note that the present discussion clearly 
shows that the space-time uncertainty relation 
is a natural generalization of modular invariance, or 
of open-closed string duality,  
exhibited by the string loop amplitudes. 

Since our derivation relies on conformal 
symmetry and is applicable to arbitrary quadrilaterals on arbitrary Riemann surfaces, which in turn can always  
be constructed 
by pasting quadrilaterals appropriately, we expect that 
the space-time uncertainty relation is robust with respect to  possible corrections to the simple setup of the present 
argument.  
In particular, the relation, being independent of the 
string coupling, is expected to be 
universally valid to all orders of 
string perturbation theory. 
Since the string coupling cannot be regarded as 
the fundamental parameter of nonperturbative 
string theory, it is natural to expect that any 
universal principle should be formulated 
independently of the string coupling. 

We have 
assumed a smooth boundary condition 
at the boundary of the rectangle. This leads to 
a saturation of the inequality of the 
uncertainty relation.  If we allow more complicated `zigzag' 
shapes for boundaries, it is not possible 
to establish such a 
simple relation as that above between 
the extremal distance and the space-time uncertainties.  However, 
we can  expect that the mean values of the space-time 
distances measured along the boundaries of 
complicated shapes in general 
increase, due to the entropy effect, 
in comparison with the case of 
smooth boundaries (namely the zero mode) 
obtained as the average of  given zigzag 
curves. 
Although there is no general proof, any reasonable definition  of the expectation value of the space-time 
distances conforms to this expectation, since the fluctuations 
contribute positively to the expectation value.  
Thus the inequality (\ref{stu}) should be 
the general expression of the reciprocity relation 
(\ref{reciprocity}) in  a direct space-time picture. 
Since the relation is  symmetric under the interchange 
$\Gamma \leftrightarrow \Gamma^*$, it is reasonable to 
suppose that the space-time uncertainty relation is 
meaningful in both limits $\Delta T \rightarrow 0$ 
and  $\Delta T \rightarrow \infty$,  
as we proposed in the previous subsection.

\section{High-energy scattering of strings and the 
space-time uncertainties}
 
We now proceed to study how the space-time uncertainty 
relation derived in the previous section is reflected in the 
high-energy behavior of string scattering. 
To the author's knowledge, a careful comparative 
investigation of the space-time uncertainty relation 
with the high-energy (and/or high-momentum transfer) behavior 
of string scattering   
has not yet been made.  We hope that the present 
section fills this gap. 

\subsection{How do we detect the interaction region from S-matix? }
In general, there are various difficulties in extracting 
precise space-time structure from on-shell S-matrix. 
This is so even in ordinary particle theories, 
since it is not possible, quantum mechanically, 
 to define the trajectories of interacting particles  unambiguously 
only from the S-matrix element. In 
string theory, the difficulties are compounded, since 
strings themselves have intrinsic extendedness. 
Therefore it is not completely clear how to  extract the space-time 
uncertainties from scattering amplitudes. 
The only conceivable way at present is to 
just treat a string state as a particle state and approximately 
trace its trajectory by forming a wave packet in space-time 
with respect to the center-of-mass coordinate. 
The effect of extendedness would then be
approximately reflected upon the 
uncertainties of the interaction region with respect to 
the center-of-mass coordinates of strings without 
referring to their internal structure.     
In our case, we have to separate the 
distance scales into temporal and spatial directions. 
We will see that the high-energy 
behavior of the scattering matrix alone does not 
allow us to carry out this completely.  But we will be able 
 to check whether the space-time uncertainty 
relation is  consistent with the high-energy behavior.

Let us consider the elastic scattering of two massless 
particles $1 + 2 \rightarrow 3 + 4$. The wave packet of 
each particle can be written as 
\EQ
\psi_i (x_i, p_i)= \int d^{9}\vec k_i \,  f_i(\vec k_i-\vec p_i)\, 
e^{i(\vec k_i \cdot \vec x_i-|\vec k_i| t_i)} ,
\EN
 where $f_i(\vec k)$ is any function with a peak at $\vec k=0$.  Here and in what follows, we assume a 10 dimensional 
flat space-time,  unless otherwise specified, neglecting the 
issue of compactification, in particular. The inverse of the 
width at the peak gives the spatial extension of the 
wave packet.    
The scattering amplitude is then given as 
\EQ
\langle 3,4| S|  1,2 \rangle = 
\Big(\prod_{i=1}^4 \int d^9 \vec k_i \Big)
f^*_3(3) f^*_4(4) f_1(1) f_2(2) \, 
\delta^{(10)}(k_1 +k_2 -k_3 - k_4) A(s, t) ,
\label{scaelement} 
\EN
where  $s=-(k_1+k_2)^2, t=
-(k_2+k_3)^2$, 
and, for brevity,  the 
momentum variables in the wave functions 
$f_i(\vec k_i -\vec p_i)$ are suppressed. 
The uncertainty of the interaction region is measured by 
examining the response of the S-matrix under appropriate shifts  
of the particle trajectories in space-time.  The wave packet,  
after given shifts $\Delta t$ and $\Delta \vec x$ of time 
and space coordinates, respectively, is 
\EQ
\psi_i (x_i, p_i ; \Delta t_i, \Delta \vec x_i)= \int d^{9}\vec k_i \,  f_i(\vec k_i-\vec p_i)\, 
e^{i(\vec k_i \cdot \vec x_i-|\vec k_i| t_i)} 
e^{i(\vec k \cdot \Delta \vec x_i-|\vec k_i| \Delta t_i)} .
\EN
To measure the uncertainty of the interaction region with respect to time, it is sufficient to choose 
$\Delta t_1=\Delta t_2=-\Delta t_3=-\Delta t_4=
\Delta t /2$ and $\Delta x_i=0$ for all $i$.  
Thus, the uncertainty $\Delta T$ can be estimated 
by examining the decay of 
the matrix element (\ref{scaelement}) under the insertion of the additional phase factor $\exp (-i(|\vec k_1|+|\vec k_2|)\Delta t) $ 
in the integrand comparing it with that without the insertion. 
On the other hand, to measure the uncertainty of the 
interaction region with respect to spatial extension, it 
is  sufficient to choose 
$\Delta \vec x_1=-\Delta \vec x_2 =\Delta \vec x^I /2$ and $\Delta t_i=0$ 
and $\Delta \vec x_3 
=-\Delta \vec x_4=\Delta \vec x^F /2$, corresponding to  the relative spatial positions of the 
trajectories of initial and final states, respectively. 
   In this case, the 
additional phase factor is 
$\exp [i(\vec k_1 -\vec k_2)\cdot \Delta \vec x^I/2 -i(\vec k_3 - \vec k_4)\cdot \Delta \vec x^F/2) ]$. 

Let us choose the center-of-mass 
system for the momenta $k_i$. 
Assuming that the scattering takes place in
 the $1$-$2$ plane and that 
the particles are all massless, we set 
\EQA
&&k_1=(-E\sin\phi/2, E\cos\phi/2, 0, \ldots, 0, E)  ,\nonumber \\
&&k_2=(E\sin\phi/2, -E\cos\phi/2, 0, \ldots, 0, E)  ,\nonumber \\
&&k_3=(E\sin\phi/2, E\cos\phi/2, 0, \ldots, 0, E) , \\
&&k_4=(-E\sin\phi/2, -E\cos\phi/2, 0, \ldots, 0, E) \nonumber  .
\EQN
Thus, 
\EQA
k_1+k_2&=&(0, 0, 0, \ldots, 0, 2E)  ,\\
k_1-k_2&=&(-2E\sin\phi/2, 2E\cos\phi/2, 0, \ldots, 0, 0) ,\\
k_3-k_4&=&(2E\sin\phi/2, 2E\cos\phi/2, 0, \ldots, 0, 0) ,\\
k_1-k_3&=&(-2E\sin\phi/2, 0, 0, \ldots, 0, 0)  ,\\
k_1-k_4&=&(0, 2E\cos\phi/2, 0, \ldots, 0, 0)  .
\EQN

The order of magnitude of the decay width with respect to 
$|\Delta t|$ is estimated by taking a small variation with respect to 
the center-of-mass energy $E$,  fixing the scattering angle $\phi$, since the variation of the additional 
 phase is just $E\Delta t$ and is independen of the angle $\phi$.   
As $\Delta t$ increases, the decay of amplitude 
becomes appreciable when the absolute value of the 
variation of the logarithm of the  
 amplitude is exceeded by the variation of the additional 
phase $E\Delta t$.  Therefore, we can roughly set 
\EQ
\Delta T\sim \langle |\Delta t| \rangle 
\sim {1\over 2}\Big|{\partial\over \partial E}
\log A(s,t) \Big| . 
\label{timeuncertainty}
\EN
This way of measuring the uncertainty should perhaps be 
regarded as giving a lower bound, since it does not 
take into account the extendedness of the initial and final string states. 
We must evaluate this quantity at the peak values of the 
momenta. Note that this expression is similar to 
the well-known Wigner formula for time delay,  
for which we usually take only the phase of the 
amplitude.   For the spreading of  the interaction region,  the variation of the modulus of  the amplitude plays an equally important role as its phase.\footnote{
When the first variation with respect to 
the integration variables 
is small, we must be careful in checking whether the 
higher variations are negligible.   For measuring 
the transverse size $|\delta x_t| $ corresponding to the shift 
of the form $\exp i(k_1-k_3) \delta x_t$, the second variation 
is indeed important. We  also note that the present 
method is reliable only when $\log A(s,t)$ is a 
smooth function. }
 
Similarly, the decay width with respect to  $|\Delta x|$ can be estimated by taking 
 variations with respect to both 
the energy and scattering angle,  
 since   the additional phase now behaves as 
$(k_1-k_2)\cdot \Delta \vec{x}^I/2-(k_3-k_4)\cdot \Delta \vec{x}^F/2=
E(-(\Delta \vec{x}^I+\Delta \vec{x}^F)_1 \sin {\phi \over 2} +(\Delta \vec{x}^I
-\Delta \vec{x}^F)_2
\cos {\phi \over 2})$, where the lower indices 
refer to the components in the 1-2 plane. 
The order of magnitude of the allowed spatial uncertainty 
is constrained by the conditions obtained by identifying the 
first variations of the modulus of the logarithm of the 
amplitude and of the additional phase
\EQ
\Big|\delta \Big(E\big(-\Delta x^{(+)}_1 \sin {\phi \over 2} +\Delta x^{(-)}_2
\cos {\phi \over 2}\big)\Big)\Big| \sim |\delta \log A|
\label{spatialbalance}
\EN
for $\Delta \vec{x}^{(\pm)}\equiv \Delta \vec{x}^I \pm \Delta \vec{x}^F$. 
This gives two equations for determining the 
components of the vector $(\Delta x^{(+)}_1, \Delta x^{(-)}_2)\equiv \Delta \vec{\tilde{x}}$ 
from the coefficients with respect to the 
variations $\delta E$ and $\delta \phi$.  
This relation shows that there are limitations  
in estimating the spatial uncertainties. First, 
since the energy variation essentially gives the 
same contribution to $|\Delta x^{(\pm)}|$ as $\Delta t$, 
 the high-energy 
scattering of massless particles can only probe 
the region in which $\Delta X \gtsim \Delta T$.  This  limitation is inevitable,  
since, for particles moving with the light velocity, a time uncertainty necessarily 
contributes to a spatial uncertainty of the same order. 
Second,  and more importantly, we can only probe the 
{\it vector} sum or difference of the spatial uncertainties for initial and final states. However, the spatial uncertainty 
for the space-time uncertainty relation should be defined to be the average of the uncertainties 
of the initial and final states as 
$\Delta X \sim (|\Delta x^I|+|\Delta x^F|)/2$. 
Due to the triangle inequality, we at least have a 
lower bound for the spatial uncertainty: 
\EQ
\Delta X > \Delta \tilde{x}. 
\label{ineq}
\EN
Note that we cannot in general expect 
this equality to be saturated, except for the very peculiar 
case where either the initial or final 
spatial uncertainty vector vanishes.  
Complete information on the space-time structure 
could only be obtained if one could completely 
convert the scattering matrix into the 
coordinate representation. Of course, for each term 
of the perturbation series, we already have such a 
picture in the form of the world-sheet 
path-integral representation. But nonperturbatively,  
in general,  we cannot expect to have such a picture.  

We remark here that the power-law behavior for high-energy fixed-angle scattering necessarily leads to 
the decrease by a factor of $1/E$ for both spatial and temporal 
uncertainties in the above sense.  This is, of course,  
typical behavior for the high-energy limit 
of local field theories.  Our task is to 
examine how string scattering deviates 
from such typical behavior of particle scattering 
in local field theories.

\subsection{High-energy and high-momentum 
transfer behavior of string scattering}

Fortunately, the behaviors of string scattering in 
the high energy limit with fixed scattering angle is 
studied in detail in Refs.\cite{grossmende} and \cite{oogurimende}.  
We study how far we can 
probe the short distance space-time 
structure using the results of these works. 
Throughout the present section, 
we use the string unit $\ell_s =1$.  

At the tree level, the leading behavior is 
\EQ
A_{{\rm tree}}(s,\phi) \sim  ig^2 2^9 s^{-1}(\sin\phi)^{-6}
\exp \Big(
{-s\over 4}f(\phi)
\Big) ,
\label{treehigh}
\EN
where 
\EQ
f(\phi)=\sin^2({\phi\over 2})\big|\log \,\sin^2({\phi\over 2})\big|
+\cos^2({\phi\over 2})\big|\log \,\cos^2({\phi\over 2})\big| .
\label{anglefunction}
\EN
Although this particular form is for bosonic closed strings, 
the main feature that the amplitude 
falls off exponentially is due only to the 
Riemannian nature of the world sheet, and hence 
the exponential 
behavior including the function (\ref{anglefunction}) 
is completely universal for any perturbative  
string theory.  

The exponential fall off of (\ref{treehigh}) 
has been regarded as one 
of the features of string theory which is 
clearly distinctive from local field theories. 
This has been the main motivation for the suggestion 
of the 
modification of Heisenberg uncertainty relation 
as (\ref{stringyuncertainty}). Indeed, if we apply the 
above method for estimating the width to the tree 
behavior directly with finite angle $\phi$, 
we would get $\Delta T \sim \Delta X \sim 
E$, corresponding to the dominant classical world sheet 
configuration (\ref{classicalworldsheet}). 
However, the exponential fall off of tree amplitudes  
only means that the tree approximation is quite 
poor for representing the high-energy behavior 
of string scattering. In fact,  for $N-1$ loop amplitudes, the exponential factor 
is replaced by $\exp (-sf(\phi)/N)$. Thus, for any small 
but finite string coupling, the high-energy limit is 
dominated by large $N$ contributions.   
The nonperturbative 
high-energy behavior is derived in Ref.\cite{oogurimende} by performing the Borel-sum over $N$. The final result there is summarized as 
\EQ
|A_{{\rm resum}}(s, \phi)| \sim \exp \big(-\sqrt{4sf(\phi)\log(1/g^2)}\,\big)
\label{borelhigh1}
\EN
for $1 \ll \log(1/g^2) \ll s \ll 1/g^{4/3}$,  
and 
\EQ
|A_{{\rm resum}}(s, \phi)| \sim \exp \big(-\sqrt{6\pi^2 sf(\phi)/\log s}\,\big)
\label{borelhigh2}
\EN
for $s\gg 1/g^{4/3}$.  The tree behavior (\ref{treehigh}) 
with a much faster decreasing exponential is valid only for 
$1 \ll s \ll \log(1/g^2)$.  

Now let us estimate the space-time uncertainties 
exhibited in the nonperturbative high-energy behavior (\ref{borelhigh2}) for fixed string coupling. 
For our purpose of estimating the 
order of magnitude for the decay width 
of the amplitudes with respect to the shift 
of the particle trajectories, we can neglect 
the imaginary part of the logarithm $\log A(s, \phi)$, 
since it only contributes to the present qualitative estimation 
at most to the same order as the real part, and hence 
it only affects the numerical multiplicative factor 
for the width. 

Using (\ref{timeuncertainty}), 
the uncertainty of the interaction region with respect to 
time is 
\EQ
\Delta T \sim \sqrt{f(\phi)} .
\label{timeuncertain}
\EN
We neglect the logarithms as well as the numerical factor 
with respect to the energy $E$, since our method 
(or any other conceivable methods) is not sufficiently 
precise to include them. 
Note that in the limit of small scattering angle, we have 
$\Delta T \sim \phi\sqrt{\log \phi} \sim (\sqrt{t}/E)\sqrt{\log(E/\sqrt{t})} \rightarrow 0$.  
 The dependence on the momentum transfer is 
strange, compared with $\Delta T \sim 1/E$   for the standard 
Regge behavior for fixed $t$.  In reality, 
the approximation used in the derivation of the 
high-energy limit will break down in the small 
angle limit, since in  that case the saddle point 
approaches a  singular boundary of the moduli space.  
Therefore we can trust our result only for 
moderate scattering angles. 

In order to obtain the uncertainty of 
the spatial interaction region, we use (\ref{spatialbalance}), 
which leads to  
\EQ
 \epsilon_1\sqrt{f(\phi)}=-\Delta \tilde{x}_1 \sin {\phi\over 2} 
+ \Delta \tilde{x}_2 \cos {\phi\over 2} ,
\label{evariation}
\EN
\EQ
\epsilon_2 {|f'({\phi})| \over \sqrt{f(\phi)}} 
=\Delta \tilde{x}_1\cos {\phi\over 2} + \Delta \tilde{x}_2 \sin
{\phi \over 2} ,
\label{phivariation}
\EN
where $\epsilon_{1}$ and $\epsilon_2$ are arbitrary sign factors, arising in 
making the comparison (\ref{spatialbalance}). 
At $\phi =\pi/2$, the first variation with respect to 
the scattering angle  
vanishes.  It is, however, easy to check that taking account of 
the second variation does not change the final conclusion 
in the limit $E\rightarrow 0$. 
 We then obtain
\EQ
4(\Delta \tilde{x})^2 \sim  f(\phi) + {f'(\phi)^2 \over f(\phi)} .
\EN
Because of the inequality (\ref{ineq}) and the relation (\ref{timeuncertain}), this gives a 
lower bound for the space-time uncertainty  relation as 
\EQ
\Delta T \Delta X > \Delta T \Delta \tilde{x} \sim 
{1\over 2}\sqrt{f(\phi)^2 +  f'(\phi)^2} 
=\sqrt{\sin^2{\phi\over 2} \Big(\log \sin{\phi\over 2}\Big)^2 
+ \cos^2{\phi\over 2} \Big(\log \cos{\phi\over 2}\Big)^2} .
\label{fixedanglestu}
\EN
 For moderate values of the scattering angle 
which are not close to 0 or $\pi$, the 
right-hand side is of order 1, independent of energy. 
This is consistent with our space-time uncertainty relation. 
In particular, this shows that we cannot probe arbitrarily 
short distances, even if both energy and momentum transfer 
increase without limit.  The fact that the 
right hand side vanishes in either limit $\phi \rightarrow 
0$ or $\phi \rightarrow\pi$ implies only that 
this  inequality  (\ref{ineq}) is far from being saturated, 
namely, $\Delta X \gg \Delta\tilde{x}$. 
For example, if we use (\ref{evariation}) and 
(\ref{phivariation}) in the limit $\phi \rightarrow 0$ of 
forward scattering, we find $\Delta \tilde{x}_2 \rightarrow \sin {\phi\over 2} 
\rightarrow 0$ and $\Delta \tilde{x}_1 \sim O(1)$ which 
indicate that the components of the spatial uncertainty 
match for the initial and final states, {\it i.e.} 
$\Delta x^I_2\sim \Delta x^F_2$, along the longitudinal direction while along the 
transverse direction there is a spread of order one. 
In any case, however, we cannot trust our formulas 
for such small scattering angle, as emphasized above.  
For a generic scattering angle, it seems reasonable to 
regard the inequality as almost saturated, namely, 
$\Delta X \sim \Delta\tilde{x}$,  since there is 
no preferred direction for the spatial uncertainty. 

\subsection{The Regge limit and space-time uncertainties}
 
We have studied the high-energy limit for fixed 
scattering angle, which means high-momentum 
transfer $s\sim t \rightarrow \infty$. 
Let us briefly consider the case of fixed momentum transfer $t=-(k_1-k_3)^2=-4E^2\sin^2{\phi\over 2}$.\footnote{
For a previous analysis of high-energy string scattering 
with fixed momentum transfer, 
see Ref.\cite{lowe} }  
Since this corresponds to the limit of small scattering 
angle, the above discussion suggests that we cannot expect 
information other than some matching 
conditions between initial and final spatial 
uncertainties. 
The high energy behavior is dominated by the 
exchange of Regge poles. In string theory, the leading 
Regge trajectory is that of the graviton. 
Hence, the tree (invariant) amplitude is given by 
\EQ
A_{{\rm tree}}(s,t) \sim g^2 {1\over t}(-is/8)^{2+t/4} .
\label{reggehigh}
\EN
 As is well known, however, 
this behavior is actually incompatible \cite{soldate} with unitarity for sufficiently high energies. 
To recover unitarity, it is again necessary to resum the 
whole perturbation series.  This problem is 
investigated in Ref. \cite{amati} using the method of 
Reggeon calculus.  It was shown that the 
series can be summed into an (operatorial) eikonal 
form in the region of 
 large impact parameter, or equivalently, 
in the region of small momentum transfer in the present momentum representation.  
In particular, the tree form (\ref{reggehigh}) 
is justified only when the eikonal 
is very small, {\it i.e.} when $1/b \sim \sqrt{t} < (g^2 s)^{-1/(D-4)} \ll 1$ is satisfied.  This is essentially the classical region.  
By reapplying the method used above to the tree amplitude 
(\ref{reggehigh}) 
in this region, 
we obtain the uncertainty in time,
\EQ
\Delta T \sim {\partial \over \partial E} \log s \sim 1/E \ll 1 .
\EN
For the uncertainty in the spatial direction, 
we can only obtain the following constraints
\EQ
|\Delta \tilde{x}_1| \sim 
{2\over E}\Big(
(1+{t\over 2}\log E)\sin{\phi\over 2} +
{\epsilon \over \sin {\phi\over 2}}(1-{t\over 2}\log E)
\Big)\sim 
{1\over \sqrt{t}}(1-{t\over 2}\log E) ,
\label{transversespread}
\EN
\EQ
|\Delta \tilde{x}_2| \sim 
{2\over E}\Big(
(1+{t\over 2}\log E)\sin{\phi\over 2} -
{\epsilon \over \sin {\phi\over 2}}(1-{t\over 2}\log E)
\Big) \sim  {4\over E} \, \, or \, \, \, {-2t\over E} \log E ,
\EN
where $\epsilon$ is the choice of relative sign between the 
energy and angle variations in extracting 
the orders of magintude for the uncertainties using (\ref{spatialbalance}). In conformity with 
the tendency found in the fixed-angle case, 
the spatial uncertainties along the longitudinal 
direction 2 match for initial and final states. 
This is expected, since the space-time 
uncertainty relation requires that the 
longitudinal spatial uncertainty 
increase with energy (or decrease of interaction time). 
The growth of the longitudinal size of a string with decreasing 
time uncertainty would be impossible 
unless the uncertainties match for 
the initial and final states along that direction. 
On the other hand, 
along the transverse directions, (\ref{transversespread}) 
indicates that the average uncertainty 
spreads without limit as the momentum transfer 
vanishes, corresponding to the exchange of a 
massless graviton.  The singular behavior of 
(\ref{transversespread}) in $t$ 
is produced by the pole at $t=0$ of the Regge amplitude. 
This is also consistent with the growth of the 
longitudinal length of strings.  From the $s$-channel viewpoint, it is very difficult to 
imagine the generation of  long-range interactions without the rapid growth of the string extension. 

That the high-energy Regge behavior of string amplitudes,  
at least  with only its simplest 2-2 elastic scattering, can only be utilized for a consistency check of the 
space-time uncertainty relation might seem somewhat disappointing. 
However, this is inevitable in view of the number of the 
 variables available in the scattering amplitude and its 
kinematical structure. 
 
We note that our method of estimating the 
interaction region directly from the high-energy behavior 
is not sensitive enough to fix the Regge intercept: 
For the property $\Delta T \sim 1/E$, only the 
power behavior with respect to energy with fixed momentum 
transfer is relevant,  and the value of intercept, 
including its sign, cannot be detected. 
Actually, in the Regge limit, this information of the Regge 
intercept, namely, the existence of massless states such as the graviton and photon in string theory, may be regarded as a consequence of the space-time uncertainty relation. 
It has long been known \cite{mueller} that
 in light-cone string theory 
there is a simple geometrical explanation for the intercept of the Regge trajectory of string theory.  We can adapt this geometrical interpretation to
 the space-time uncertainty relation as follows. 

Consider the elastic scattering of two strings, 
$p_1 + p_2 \rightarrow p_3 + p_4$ in the extreme forward 
region, where the longitudinal momenta $p^+_1$ and $p^+_3$ 
are very large compared with others.  Namely, we choose a sort of a laboratory frame instead of the center-of-mass frame. 
If we treat the high-momentum state as the target string and 
the low-momentum state as the projectile string, it is 
natural to represent the interaction by the insertion of the 
vertex operators corresponding to the absorption and emission 
of the projectile string onto the target string state. 
In this case, the projectile string can effectively be 
treated as a probe with small longitudinal extension, 
since the momentum associated with the vertex operator is 
small. On the other hand, by reversing the roles of projectile and target strings we see that the intermediate state 
induced by the interaction has a large longitudinal 
extension. Also,  by repeating the above 
analysis of the Regge limit in the present frame,  we can see that the interaction time 
is small and the longitudinal extension associated with 
initial and final states must match  each other in the Regge limit.   Note that the main difference between this situation 
and that in the center-of-mass frame is only that 
$s\sim p_1^+p_2^-$ instead of $s\sim E^2$. This means that 
 the probability 
for the interaction with forward scattering is proportional to its longitudinal length,  which can be regarded as being 
proportional to the longitudinal momenta, 
since the interactions of strings are regarded as occurring  independently  
at each segment of the target string.\footnote{
Here it is important that the string is a continuous object. 
If, for example, we consider some extended object with 
only a discrete and finite number of 
degrees of freedom, we cannot expect to 
generate a graviton or any massless particles naturally. 
It seems very difficult to construct a reasonable theoretical framework 
other than string theory that 
contains gravity and satisfies the space-time 
uncertainty relation. 
}
With the identification of the longitudinal length and the 
longitudinal momentum in accordance with the 
space-time uncertainty relation, this means that the 
probability amplitude in the tree approximation linearly grows 
with large longitudinal momentum $p^+_1
\sim p^+_3$. For the invariant amplitude, this amounts to 
the Regge intercept $\alpha(0) = 2$.  If we only consider the 
open string interactions, neglecting the closed string, 
the same argument leads to $\alpha(0)=1$, since the 
interaction only occurs at the string end point and hence the 
probability amplitude is constant in the high-energy limit. 

\subsection{A remark: minimum nonlocality ?} 

Finally we remark that there is no 
guarantee that the Borel summation of the leading 
behavior of the perturbation series gives a 
unique nonperturbative result.  
Therefore, the formula which we have relied upon 
for studying the fixed angle scattering may not be 
completely correct, due to some 
nonperturbative effects that have not been 
taken into account in the Borel summation. 

However, 
at least for a certain finite range of the string coupling 
including the weak coupling regime, it seems reasonable 
to regard the properties found here 
as providing evidence for the following viewpoint: 
The space-time uncertainty relation 
is a natural principle which characterizes string theory 
nonperturbatively  
as being {\it minimally but critically} departed 
from the usual framework of local 
field theory for resolving  ultraviolet difficulties.  
This view may be supported 
by recalling that the high-energy behavior (\ref{borelhigh2}) 
with fixed scattering angle almost realizes the 
fastest allowed decrease of the form, \cite{celmar} 
$ \exp( -f(\phi)\sqrt{s}\log s)$.  
The proof of this theorem uses, apart from the usual analyticity and unitarity, assumptions of 
polynomial boundedness in the energy for fixed $t$ and 
also of the existence of a mass gap.  Obviously, the latter is 
not satisfied in the presence of graviton.  
However, that this lower bound just represents 
the behavior, up to 
logarithms,  
corresponding to the saturation of the bound expressed in the  
space-time uncertainty relation, as is exhibited 
by (\ref{fixedanglestu}), is very suggestive.  
We may say that `locality' is almost satisfied in some effective 
sense in string theory 
from the viewpoint of the analyticity property 
of scattering amplitudes.\footnote{In the literature, we can find  another approach to locality based on the 
commutation relation of string fields. \cite{martinec}$^,$\cite{polsusslowe}
It would be an interesting problem to  connect the latter approach to our space-time uncertainty relation. } 
The space-time uncertainty relation may be interpreted as 
the basic principle for introducing nonlocality 
in a way that does not contradict the analyticity 
property of the scattering amplitude,  whose validity is usually 
assumed  for local field theories.  

\section{The meaning of space-time uncertainty relation} 
 
Now that we have checked the consistency of the space-time uncertainty relation with high-energy string scattering, let us study its implications  
from a more general standpoint. Since the 
relation expresses a particular way by which  string theory 
deals with the short distance structure of space-time, we 
expect that it should predict (or explain) some characteristic features 
of string theory, when combined with other physical 
characteristics of the theory.  

\subsection{The characteristic scale for microscopic black holes in string theory}
We first consider an 
implication for microscopic gravitational phenomenon. 
Usually, the characteristic scale of quantum gravity 
is assumed to be the Planck scale, which in ten-dimensional 
string theory is equal to $\ell_P \sim g_s^{1/4} \ell_s$,  corresponding to the ten-dimensional Newton constant 
$G_{10}\sim g_s^2\ell_s^8$. 
Indeed, if we neglect the effect of higher massive modes of string theory, 
this would be the only relevant scale.   
 Let us consider the 
limitation of the notion of classical space-time 
from this viewpoint in light of the possible formation 
of black holes in the short distance regime. 
Suppose that we  probe 
the space-time structure at a small 
resolution of order $\delta T$ along the 
time direction. This necessarily induces an uncertainty 
$\delta E \sim 1/\delta T$ of energy.  Let us require 
ordinary flat space-time structure to be qualitatively preserved 
at the microscopic level 
by demanding that 
 no virtual horizon is encountered, associated with this uncertainty 
of energy.  Then we have to impose the condition 
 that the minimum resolution along spatial directions must be 
larger than the  Schwarzschild radius corresponding to this energy:
\[
\delta X \gtsim (G_{10}/\delta T )^{1/7}. 
\]
This leads to the `black-hole uncertainty relation' \footnote{
Similar  relations have been considered  by other authors,  
independently of string theory. However our interpretation is somewhat different from those of other works. (See for a recent example Ref.\cite{bhuncertain})  We also note that the power $1/7 \, (=1/(D-3))$ in the right hand side depends on the space-time dimensions. In particular, for $D=4$ the left-hand side of the black hole uncertainty relation takes the same form as 
the stringy one (\ref{stu}). In connection with this, see an 
interesting paper.\cite{thooft2} The author would like to 
thank M. Li for bringing this last reference to his attention. }
\EQ
\delta T (\delta X)^7 \gtsim G_{10}. 
\label{gravitationaluncertainty}
\EN
This  expresses a 
limitation, for observers  
at asymptotic infinity,  with respect to 
 spatial and temporal resolutions,  below which 
the naive classical  space-time picture without 
the formation of microscopic black holes can no longer be applied. 
If we assume that the spatial and temporal scales are of the 
same order, this would lead to the familiar looking relation 
$
\delta T\sim \delta X \gtsim \ell_P. 
$
However, in the presence of some stable very massive 
particle state in probing the short distance scales,  such as a D-particle, 
this assumption 
may not necessarily be valid, and we should in general 
treat the two 
scales independently.  

Furthermore, 
 it is important to remember that 
the relation (\ref{gravitationaluncertainty}) 
does {\it not} forbid 
{\it smaller} spatial scales than $\delta X$ {\it in principle}. 
Suppose we use as a probe 
 a sufficiently heavy particle, such as a D-particle 
in the weak string-coupling regime.  We can then neglect the extendedness of the 
wave function  and localize the particle in an 
arbitrarily small region.  In this limit, classical 
general relativity can be a good approximation. 
But general relativity 
only requires the existence of 
local time, and hence we cannot forbid the formation of  black holes. 
This only stipulates that we cannot read the clock on the 
particle inside the black hole from an asymptotic region at infinity. 
If we suppose a local observer (namely just another particle) 
sitting somewhere apart 
in a local frame which falls into the black hole, 
it is still meaningful 
to consider the local space-time structure at scales which 
exceed the condition (\ref{gravitationaluncertainty}), since the extendedness of the  wave packet of a sufficiently heavy particle 
can, in principle, be less than the limitation set by 
(\ref{gravitationaluncertainty}).  In connection 
with this, it should be kept in mind that the above condition 
only corresponds to the restriction on 
 the formation of microscopic 
black holes. For example, for a light probe,  instead of 
a very heavy one, we have to take into account the usual 
quantum mechanical spread of the wave function,   
 as we will do below in deriving the characteristic scale 
of D-particle scattering. 

Despite a similarity in its appearance to (\ref{gravitationaluncertainty}),   
the space-time uncertainty relation 
of full string theory places a limitation in principle 
on the scale  beyond which 
we can never probe the space-time structure by any 
experiment allowed in string theory;
\EQ
\Delta T \Delta X \gtsim \ell_s^2 .
\label{stu2}
\EN
Note that such a strong statement is 
acceptable in string theory, because it is a well-defined theory 
resolving the ultraviolet problems. The 
nature of the condition (\ref{gravitationaluncertainty}) 
is therefore quite different from (\ref{stu2}). 
In this situation, 
the most important scale corresponding to truly 
stringy phenomena is where 
these two limitations of different kinds meet.   
Namely, beyond this crossover point, it becomes 
completely meaningless to talk about the classical geometry of 
a black hole, and hence it is where the  true limitation 
on the validity of classical general relativity must be set. 
The critical scales $\Delta T_c$ 
and $\Delta X_c$ corresponding to 
the crossover are obtained by substituting 
the relation $\Delta T_c \sim \ell_s^2/\Delta X_c$ into 
(\ref{gravitationaluncertainty}):
\EQ
(\Delta X_c)^6 \sim {G_{10}\over \ell_s^2}=g_s^2 \ell_s^6 .
\EN
This gives 
\EQ
\Delta X_c \sim g_s^{1/3}\ell_s , \quad 
\Delta T_c \sim g_s^{-1/3}\ell_s. 
\EN
 Interestingly enough, we have derived the well-known 
eleven dimensional M-theory scale 
\EQ
\ell_M = g_s^{1/3}\ell_s =\Delta X_c
\EN
as the critical spatial scale, without invoking 
D-branes and string dualities directly.  
Note that this critical scale crucially depends on 
10 dimensional space-time.  For example, in 4 space-time 
dimensions, there is no such critical scale for 
arbitrary values of string coupling: Namely, there is only a `critical coupling' $g_s \sim 1$ at which the Planck scale and string scale coincide.   
\begin{center}
\begin{figure}
\begin{picture}(240,220)
\put(80,0){\epsfxsize 230pt  \epsfbox{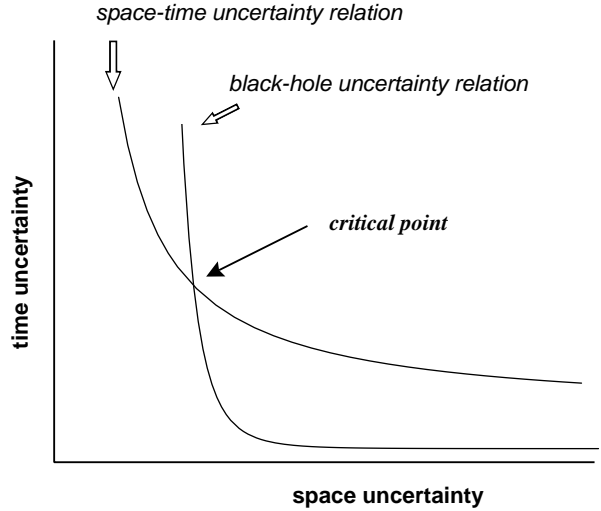}}
 \end{picture}
\caption{This diagram  
schematically shows the structure of the 
space-time uncertainty relation and the black hole  
uncertainty relation. The critical point is where the 
two relations meet. }
\label{fig1} 
\end{figure}
\end{center}
To appreciate the meaning of the critical scales, it is 
useful to look at the diagram in Fig. \ref{fig1}.  
We see clearly that  for $\Delta t < \Delta T_c$ 
there is no region where the concept 
of the microscopic black hole associated with 
quantum fluctuations is meaningful. 
On the other hand, in the region $\Delta t > \Delta T_c$, 
there is a small region where $(\Delta t)^{-1}\ell_s^2 < \Delta X < \Delta X_c$ 
is satisfied, and 
hence  black hole formation at the 
microscopic level may be meaningful in string theory. 
The importance of this region increases as the string 
coupling grows larger.  In the limit of weak string coupling,  where 
$\Delta T_c \rightarrow \infty $ and $ \Delta X_c\rightarrow 0$, 
there is essentially no  fluctuation of the space-time metric 
corresponding to the formation of microscopic black holes.  
Unfortunately, the 
space-time uncertainty relation alone cannot 
predict more detailed properties of stringy black holes 
at microscopic scales. It is an important 
problem to explore the physics in this region 
in string theory. The above relation 
between the space-time 
uncertainty relation and the black hole 
uncertainty relation suggests 
that in the strong string-coupling regime and in the region $\Delta T > \Delta T_c (\ll \ell_s)$, the black-hole 
uncertainty relation essentially governs the 
physics at long distatances $\Delta X > R_S$, with 
$R_S$ being the Schwarzschild radius, since  
$R_S > \ell_s^2 /\Delta T$ with $ \Delta T \sim 1/E$ 
there.  

\subsection{The characteristic scale of D-particle dynamics}

In the case of  high-energy string scattering, 
we could not probe the region $\Delta X < \Delta T$. 
To overcome this barrier, we need massive 
stable particles.  The point-like D-brane, {\it i.e} a D-particle, of 
the type IIA superstring theory is an ideal 
agent in this context, at least for a sufficiently weak string coupling,  
since its mass is proportional to $1/g_s$ and 
its stability is guaranteed by the BPS property. 
The derivation of the characteristic scale of D-particle interactions
 has been 
given in two previous works.\cite{liyo}\cite{liyo2} However, for the purpose of 
selfcontainedness and for comparison with the result of the 
previous subsection, 
we repeat the argument here with some 
clarifications.  

Suppose that the region we are trying to probe by the scattering of two D-particles 
is of order $\Delta X$. 
Since the characteristic spatial extension of open strings 
mediating the D-particles is then of 
 order $\Delta X$, we can
use the space-time uncertainty relation. 
The space-time uncertainty relation 
demands that the characteristic velocity $v$ of D-particles    
is constrained by 
\[
\Delta T \Delta X \sim {(\Delta X)^2 \over v}
 \gtsim \ell_s^2 , 
\]
since the period of  time required for the experiment is 
of order $\Delta T \sim \Delta X/v$. 
Note that the last relation is 
due to the fact that $\Delta T$
 is the time interval during which the length of the open 
string is of order $\Delta X$. 
This gives the order of the magnitude for the  minimum possible
distance probed by D-particle scattering with
velocity $v$:
\EQ
\Delta X\gtsim 
\sqrt{v}\ell_s.
\label{velocity-distance}
\EN
Thus to probe spatial
distances shorter 
than the time scale, {\it i.e.} $\Delta X \ll \Delta T$, we have to use D-particles with small
velocity $v \ll 1$.  However,  the spreading of the wave packet increases with decreasing velocity as 
\EQ
\Delta X_w \sim
\Delta T\Delta_w v \sim
{g_s\over v}\ell_s ,
\EN
since the ordinary time-energy uncertainty relation
asserts that the uncertainly of the velocity is of
order $\Delta_w v\sim g_sv^{-1/2}$ for  time
interval of order $\Delta T \sim v^{-1/2}\ell_s$. 
To probe a range of spatial distance $\Delta X$, 
we must have $\Delta X \gtsim \Delta X_w$. 
Combining these two conditions, we see that the
shortest spatial length is given by
\EQ
\Delta X \sim  g_s^{1/3}\ell_s ,
\label{plancklength}
\EN
and the associated time scale is
\EQ
\Delta T \sim  g_s^{-1/3}\ell_s .
\label{timescale}
\EN
Thus the minimal scales of D-particle-D-particle 
scattering coincide with the critical scales 
for microscopic black holes derived above. 
In other words,  the minimal scales 
of D-particle scattering is just characterized by the 
condition that the fluctuation of the metric 
induced by the D-particle scattering is automatically restricted 
so that no microscopic 
black holes are formed during a scattering process.   
Indeed, we can derive the same scales from the black-hole 
uncertainty relation (\ref{gravitationaluncertainty}) by 
using the restriction $\delta T/m\delta X\sim \delta X$ for 
the spreading of the wave packet of a free particle with mass $m
\sim 1/g_s\ell_s$ which is localized within a spatial uncertainty of order $\delta X$, conforming to the above agreement. 

In view of this interpretation of the scale of D-particle dynamics,  
the agreement between D-particle scales and those 
for microscopic black hole formation is consistent with  the seemingly 
surprising fact that 
the effective supersymmetric Yang-Mills  quantum mechanics,  
  which is formulated on a flat Minkowski background and  does not, at least manifestly, have any 
degrees of freedom of the gravitational field,  
can reproduce \cite{dkps}$^{\sim}$\cite{okyo} the gravitational interaction of type IIA supergravity, or 
equivalently, of the 11 dimensional supergravity with 
vanishingly small compactification radius $R_{11}=g_s\ell_s$, 
in the weak string-coupling (perturbative) regime. 
Naively, we expect that the 
supergravity approximation to string theory 
is only valid at scales which are larger than the 
string scale $\ell_s$.  On the other hand,  
the Yang-Mills approximation, keeping only the 
lowest string modes, is in general regarded as being 
reliable in the 
regime where the lengths of open strings connecting 
D-particles are small compared with the string 
scale. However, the consideration given in  the 
previous subsection indicates that truly 
stringy gravitational phenomena are characterized by the 
critical scales $\Delta T_s\sim g_s^{-1/3}\ell_s\gg \ell_s, \Delta X_c 
\sim g_s^{1/3}\ell_s \ll \ell_s$.  
Given the fact that the Yang-Mills approximation to string theory 
 is characterized  precisely by the same scales, the compatibility of Yang-Mills approximation with  supergravity 
can naturally be accepted as a consistency check of our chain 
of ideas at a `phenomenological' level.  

It should be kept in mind that 
 the present discussion is certainly not  sufficient for explaining 
the agreement of the Yang-Mills description with 
supergravity in the long distance regime.  Why such Yang-Mills 
models can simulate gravity is still largely in the realm of mystery, 
since Yang-Mills theory has no symmetry 
corresponding to general coordinate transformations, 
 and also since 
it has no manifest Lorentz invariance, either as an 
effective 10-D theory or as an infinite-momentum frame 
description of 11-D theory following the Matrix-theory conjecture.    
At least in the lowest order one-loop 
approximation,\cite{dkps} the agreement is 
explained by the constraint coming from supersymmetry.  
It seems hard to believe, however, 
that  global supersymmetry {\it alone}  can explain the quantitative agreement of 3-body interactions found 
in Ref.\cite{okyo} which is a genuinely nonlinear effect 
of supergravity.  But this might turn out to be an incorrect prejudice. 
 For a recent detailed discussion on the role of 
supersymmetry in general Yang-Mills matrix models, see Ref.\cite{kazamamura} and references cited therein.   

As the next topic of this subsection, we consider the 
D-particle scales from a slightly different viewpoint of 
degrees-of-freedom counting. 
Although the space-time uncertainty relation is 
first derived by a reinterpretation of the ordinary 
quantum mechanical uncertainty relation between 
energy and time, the fact that it puts a limitation on the 
observable length scales suggests that 
it should also imply a drastic modification on the counting 
of physical degrees of freedom.  
Let us consider how it affects the quantum state itself 
in the case of D-particles.  The discussion of the previous 
subsection on D-particle scales emphasized the 
possible scale probed by the dynamical process of scattering.  
It is not obvious whether the same scale is relevant 
for restricting the general quantum state.  The following 
derivation of the scale suggests that the same scale indeed 
is important from this viewpoint too. 

Consider the state of a D-particle which is localized within a 
spatial uncertainty $\Delta X$. 
The ordinary Heisenberg
relation $\Delta p \Delta X \gtsim 1$, which 
is the usual restriction on the degrees of freedom 
in quantum theory,  then 
gives the relation for the velocity
\[
v \gtsim {g_s\ell_s \over \Delta X} .
\]
On the other hand, the space-time uncertain relation,  
reflecting the interaction of D-particles 
through open strings,  
implies the condition (\ref{velocity-distance}) for the minimum  meaningful 
distances among D-particles with given velocities of order 
$v$ as   
\[
\Delta X \gtsim \sqrt{v}\ell_s .
\]
By eliminating the velocity, we obtain
 the same condition on the scale 
of localization  $\Delta X \gtsim g_s^{1/3}\ell_s$ 
of a D-particle state at a given instant of time.  
 In the M-theory interpretation 
of the D-particle, this is consistent with the holographic 
behavior that the minimum bit of quantum information stored in a D-particle state 
is identified with the unit cell whose volume in the 
transverse dimensions is of the order 
of the 11-dimensional Planck volume $\ell_{11}^9 \sim g_s^3 \ell_s^9 \sim (\Delta X)^9$.

Finally, we explain how these characteristic scales of D-particle dynamics 
are embodied in the 
effective Yang-Mills quantum mechanics: It can best be 
formulated by a symmetry property, called `generalized  
conformal symmetry',  which is proposed in the 
Ref.\cite{jeyo} and further developed in Ref.\cite{jekayo}$^{\sim}$\cite{yo4}.   
Briefly, the effective action, 
suppressing the fermionic part, 
\EQ
S =\int dt \, \Tr
\Bigl( {1\over 2g_s\ell_s} D_t X_i D_t X_i + i \theta D_t \theta
+{1 \over 4g_s\ell_s^5} [X_i, X_j]^2 -
 .... ) 
\label{action}
\EN
of the supersymmetric 
Yang-Mills matrix quantum mechanics 
is invariant under the transformations 
\EQ
X_i \rightarrow \lambda X_i, \, \,  \,  t\rightarrow \lambda^{-1}t, \, \, \, 
g_s\rightarrow \lambda^3 g_s ,
\label{scaling}
\EN
\EQ
\delta_K X_i = 2 \epsilon  t X_i , \, \, \, 
 \delta_K t =-  \epsilon t^2 , \, \, \,  \delta_K g_s =6\epsilon  t g_s , 
\label{eq29}
\EN 
which together with the trivial time translation symmetry form an 
$SO(2,1)$ algebra. This shows that the 
characteristic scales of the theory are indeed 
(\ref{plancklength}) and (\ref{timescale}). 
Combining this with the fact that the same symmetry is 
satisfied in the classical metric of the D0 solution 
of type-IIA supergravity and with the  help of some constraints 
due to supersymmetry, it is demonstrated in Ref.\cite{jeyo} that the generalized conformal symmetry 
can determine the effective D0 action as a probe 
to all orders in the velocity expansion, within the 
eikonal approximation, neglecting time derivatives of the 
velocity. 

An important point here is that 
the supersymmetry of the model plays a crucial role for ensuring that 
the D-particles can be free when the distances among them 
are sufficiently large. 
Without the supersymmetry we would have 
nonvanishing zero-point energies. The zero-point energies 
contribute to the effective static potential, which grows linearly with distances.  This would render  scattering experiment 
impossible.  If we assume the scaling symmetry (\ref{scaling}), 
 the effective action 
for two-body scattering  in general takes the form 
\EQ
S_{eff}=\int dt \Bigl(
{1\over 2g_s\ell_s}v^2 -\sum_{p=0}^{\infty} c_p {v^{2p}\ell_s^{4p-2}
\over r^{4p-1}}
+O(g_s)
\Bigr) 
\EN
in the limit of weak coupling 
 with $c_p$ representing  numerical constants.  
The zero-point oscillation 
corresponds to the first term, $p=0$. It is well known that 
supersymmetry eliminates the next term, $p=1$, too, 
and the effective interaction starts from the $p=2$ term,  $v^4/r^7$. 

As a further remark, we note that  the product 
$\delta X \delta t$ of small  variations is invariant 
under the above transformations, suggesting that 
the generalized conformal symmetry may be a part of 
a more general set of transformations which characterize 
the algebraic structure associated with 
the space-time uncertainty relations. 
Just as the canonical structure of classical phase space 
is transformed into Hilbert space of physical states 
in quantum theory,  
which is characterized by the `unitary structure', 
such a characterization might lead to some 
appropriate mathematical structure 
underlying  the space-time uncertainty relation. 
Exploration of such ideas 
might be an important future direction. 
However, this issue will not be addressed in the present paper.  
To carry out such a study meaningfully, we need more data. 

For example, the Yang-Mills models of above type 
cannot describe the system with both D-branes and 
anti D-branes.  Once the D-branes and anti D-branes are 
both included,\cite{sussbanks} we have no justification for the approximation 
retaining only the lowest string modes, as the following 
argument shows.    
In the simplest approximation in which we 
retain only the usual gravitational interaction, 
the effective action is 
\[
\int dt \Bigl(
{1 \over 2g_s\ell_s}v^2+{\ell_s^6\over r^7}
\Bigr)
\]
If we assume that the space-time uncertainty relation 
is saturated at its lower bound, the relation $r^2 \sim v\ell_s^2$ leads to 
an estimate of the characteristic length 
scale as $r_c \sim g_s^{1/11}\ell_s$ which is 
smaller than the string scale $\ell_s$ in the weak coupling 
region, while it is somewhat larger than the critical spatial 
scale of D-particle-D-particle scattering. 
Since, however, the string scale $\ell_s$ is 
just the characteristic scale corresponding to the instability, 
we have to take into account tachyons, and all higher modes too which are characterized by the same string scale, 
in terms of open strings.  In the case of 
pure D-particle systems, the validity of retaining 
only the lowest open string modes and consistency with 
supergravity at least in the lowest order 
approximation in the weak string coupling is ensured by the 
supersymmetry: It leads to the fact  that 
both short-distance and long-distance forces are 
described by the lowest Yang-Mills modes alone 
 in the approximation of 
one-graviton exchange.  Without manifest supersymmetry, however, there is no such mechanism which may ensure 
the validity of a field theory approximation. 

A conclusion of this simple argument is that 
the D-particle and anti-D-particle system cannot be 
assumed to saturate the lower bound of the space-time uncertainty relation. 
In fact, if we just apply the ordinary Heisenberg uncertainty relation 
for the Hamiltonian $H={g_s \ell_s p^2 \over 2}
-{\ell_s^6\over r^7}$, we get a much larger spatial scale 
of order $\Delta X \sim g_s^{-1/5}\ell_s\gg \ell_s$. 
If we further assume that the scattering occurs through a 
metastable resonant state, the characteristic time scale is 
$\Delta T \sim g_s^{-7/5}\ell_s$, which leads 
to $\Delta X\Delta T \sim g_s^{-8/5}\ell_s^2\gg \ell_s^2$. 
It seems that the lower bound of the space-time 
uncertainty relation is expected only for 
some particularly symmetric systems, such as 
systems satisfying the BPS condition and 
the generalized conformal symmetry.  
This expectation is also in accord with the results of 
high-energy fixed angle (or high-momentum transfer) 
 scattering of strings obtained in the previous seciton, 
if one supposes that through the high-energy fixed angle 
scattering we are probing a regime where 
the symmetry is much enhanced.

At this juncture, it is perhaps worth remarking also that the
generalized conformal symmetry
is regarded as the underlying symmetry for the so-called 
DLCQ interpretation of the Yang-Mills matrix model.  
We can freely change the
engineering scales in analyzing the system. 
Thus, if we wish to
keep the numerical value of the transverse
dimensions, we perform the rescaling $t\rightarrow \lambda^{-1} t,
X_i\rightarrow \lambda^{-1}X_i ,\ell_s\rightarrow
\lambda^{-1}\ell_s $ simultaneously with the
generalized scaling transformation leading to
the scaling
$t\rightarrow \lambda^{-2} t, \, \, 
  X_i\rightarrow X_i, \, \, R\rightarrow \lambda^2 R$ and $\ell_{11}\rightarrow \ell_{11}$, 
which can be interpreted  as a kinematical boost
transformation along the 11th direction 
compactified with radius $R_{11}=g_s\ell_s$. Alternatively,
we can keep the numerical value of
time or energy by making the rescaling
$t\rightarrow \lambda t,\quad X_i\rightarrow \lambda X_i,
\, \ell_s \rightarrow \lambda \ell_s$,
leading to the scaling
$t\rightarrow t, \, \, 
R\rightarrow \lambda^4 R, X_i \rightarrow \lambda^2 X_i,\, \, 
\ell_{11}
\rightarrow \lambda^2\ell_{11}$ and $\ell_s \rightarrow \lambda \ell_s$, which is in fact equivalent to the
`tilde' transformation utilized in Ref.\cite{seibergsen} in an 
attempt to justify the 
Matrix theory for finite $N$.   
Note that
although the second case makes the
string length $\ell_s$ small by assuming small $\lambda$,
the length scale for transverse directions smaller than the
string scale is always  even smaller 
$(<\lambda^2\ell_s)$.   For further discussions and applications of the generalized conformal symmetry in D-brane dynamics, we refer the reader to our previous
 papers cited above. Here we only mention that the 
generalized conformal symmetry provides a basis for the 
extension of the AdS/CFT correspondence for the Yang-Mills matrix model.  The concrete computation of the correlators 
 led to somewhat unexpected but suggestive 
results with regard to the question of 
the compatibility of Lorentz invariance and holography 
in Matrix-theory conjecture, as fully discussed in Refs.\cite{seyo} and \cite{yo4}.

\subsection{Interpretation of 
black-hole complementarity and UV-IR correspondence}

In the first part of the present section, we have 
emphasized the relevance of  the space-time uncertainty relation 
to the question of formation of microscopic black holes through 
 the fluctuation of space-time geometry.  
Is this relation also relevant  for macroscopic black holes? 
Qualitatively at least, one thing is clear: 
For an external observer sitting outside black holes, 
strings are seen to more and more spread as they approach to 
the horizon, because of the infinite 
time delay near the horizon.  
Namely, for the observer far from the 
horizon, the uncertainty of time  
becomes small, $\Delta T \rightarrow 0$, without limit 
as strings approach to the horizon. 
The space-time uncertainty relation then demands 
that the spatial uncertainty increases as $\Delta X
\sim \ell_s^2 /\Delta T \rightarrow \infty$ 
without limit. This phenomenon is the basis for the proposal 
of implementing the principle of `black-hole complementarity'  
 \cite{sussetal} 
in terms of string theory by Susskind \cite{suss2} in 1993. 
The general space-time uncertainty relation (\ref{stu}) 
proposed earlier just conforms to this principle of black-hole physics.   In fact, a version of the space-time uncertainty relation is  
rederived in light-cone string theory 
in Ref.\cite{suss2} from the viewpoint of black-hole complementarity. 

However, starting from microscopic string theory, 
it is in general an extremely difficult dynamical problem  to deal with macroscopic black holes involving 
string interactions in essential ways and resulting in 
macroscopic scales quite different from the fundamental 
string scale.  Thus we cannot be 
completely sure in identifying
 the concrete physical consequences of the 
above general property of strings near the horizon.   
In the present subsection, we give a reinterpretation of  the Beckenstein-Hawking 
entropy of macroscopic black holes from the viewpoint 
of the space-time uncertainty relation following the general 
idea of black-hope complementarity.  
Although most of what we discuss here may 
simply represent different ways 
of expressing points which have been discussed 
previously, we hope that our presentation 
at least has the merit of looking at important things from 
a new angle. 

As already alluded to in our derivation of the 
space-time uncertainty relation, one of the 
crucial properties of a free string, which 
is responsible for the space-time uncertainty principle, 
is its large degeneracy 
[$d(E) \sim \exp k\ell_s E$] as energy increases. 
It is reasonable to suppose that this property is 
not qualitatively spoiled by the interaction of strings, which 
must definitely be taken into account for the treatment 
of macroscopic phenomenon.  

Based on this expectation, our fundamental 
assumption is that the entropy of macroscopic Schwarzschild black hole  
is  given by 
\EQ
S = \log W  \sim \Delta X_{{\rm eff}}/\ell_s
\label{entropy1}
\EN
 where   $\Delta X_{{\rm eff}}$ is 
the {\it effective}  spatial uncertainty of the state. 
The space-time uncertainty relation then leads to 
a lower bound  in terms of the effective   
uncertainty $\Delta T_{{\rm eff}}$ along the time direction as
\EQ
S \gtsim \ell_s/ \Delta T_{{\rm eff}}. 
\label{entropy2}
\EN
 Intuitively, the motivation for 
this proposal should be clear: We have  replaced the 
energy by the uncertainties in the formula of the 
degeneracy [$W\sim d(E)$] of a free string state.  
In particular, the form $\Delta X_{{\rm eff}}/\ell_s$ is natural 
if we assume  that the macroscopic state is 
effectively described as a single string 
state with effective longitudinal length $\Delta X_{{\rm eff}}$ corresponding to 
the effective spatial uncertainty.  The assumption 
that near a black hole horizon the state should be 
treated as a single string state seems natural 
in view of the exponentially large degeneracy, as 
previously argued, e.g. in Ref.\cite{suss2}.  

The effective uncertainties in general should depend on how precisely 
the states are specified.  That a state is  macroscopic means 
that it is specified solely by the macroscopic 
variables of state, such as the mass, temperature, total 
angular momentum, and so on.  
In the case of a Schwarzschild black hole, such macroscopic 
parameters are only its mass $M$ and its Schwarzschild 
radius $R_S$. We treat these two parameters 
as being independent, since the gravitational 
constant is regarded as an independent dynamical 
parameter corresponding to the vacuum expectation value of the dilaton. 
Now, on dimensional grounds, the effective spatial uncertainty 
must take the form 
\EQ
\Delta X_{{\rm eff}}=\ell_s f\Big({R_S\over \ell_s}, M\ell_s\Big) .
\label{conversion1}
\EN
However, the entropy of a macroscopic state should be 
expressible only in terms of macroscopic 
parameters, the function $f({R_S\over \ell_s}, M\ell_s)$ 
 actually depends only on the product of the 
variables without explicit dependence of the string length $\ell_s$: 
\[
f({R_S\over \ell_s}, M\ell_s) =f(R_sM)  .
\] 
To fix the form of the function $f(x)$ of a single variable, 
we here invoke the 
`correspondence principle' \cite{suss2}$^,$\cite{horopol} that the 
black hole entropy must be reduced to 
$\log d(M)\sim \ell_s M$ at the point where the Schwarzschild 
radius becomes equal to the string scale $R_S$, namely in the limit $R_S\rightarrow \ell_s$. This 
immediately leads to $f(x) \sim x$. Thus we obtain
 the entropy of the Beckenstein form in $D$ dimensional space-time, 
\EQ
S \sim R_SM \sim (G_D M)^{1/(D-3)}M \sim G_D^{-1}
(G_DM)^{(D-2)/(D-3)} ,
\EN
where $G_D$ is the Newton constant in $D$-dimensions, 
$G_D\sim g_s^2\ell_s^{D-2}$. 

The characteristic effective time uncertainty 
$\Delta T\sim \ell_s/(R_SM)$ associated with this reinterpretation 
of the black hole entropy can be understood from the 
viewpoint of `stretched horizon' which is assumed 
to be located at a distance of order $\ell_s$. 
As is well known, the near horizon geometry 
of a large Schwarzschild black hole 
is approximated by the Rindler metric 
$ds^2 = -\rho^2 d\tau_R^2 + d\rho^2 + ds_{{\rm transverse}}^2$  
whose time $\tau_R$ is related to the 
Schwarzschild time (namely the time which is 
synchronized with a clock at infinity) by 
$\tau_R \sim t/R_S$.  The time scale 
at the stretched horizon $\rho \sim \ell_s$ must be scaled by 
$\ell_s$.  Then, a Schwarzschild time scale 
of order $1/M$ is converted to a proper 
time scale $\ell_s/R_SM$ at the stretched horizon. 
Thus the {\it effective} uncertainties are 
essentially the uncertainties at the stretched horizon 
measured in the 
Rindler frame \cite{suss2} describing the near-horizon geometry 
of a macroscopic black hole. 

Our arguments, though admittedly mostly the consequences of 
simple dimension counting and hence yet too crude,  seem to show the basic conformity of the space-time uncertainty principle with  
 black hole entropy, and perhaps with the 
property of holography,\cite{thooft}$^,$\cite{suss3}  which is expected to be 
satisfied in any well-defined quantum theory of gravity. 
The information of a macroscopic black hole is encoded within the 
spatial uncertainty of order $\Delta X_{{\rm eff}}\sim R_SM\ell_s$. Or in terms of time, 
this corresponds to the effective time resolution 
of order $\Delta T_{{\rm eff}} \sim \ell_s/(R_SM)$ at the lower bound for 
the entropy. 
At first sight, the last relation
 may seem quite counter intuitive, since it 
suggests a time scale much smaller than the string scale 
for understanding a macroscopic object. 
But it is not so surprising if we recall that 
this is precisely where the black-hole horizon plays the role 
as the agent for producing an infinite delay with respect to time duration. 
Although the horizon is not singular at all in terms of classical 
{\it local} geometry,  it plays a quite singular role 
in terms of quantum theory, which {\it cannot} be formulated 
in terms of local geometry alone because of the 
superposition principle. This is one 
of the fundamental conflicts between general relativity and 
quantum theory, from a conceptual viewpoint. 
The space-time uncertainty relation demands that 
this conflict should be resolved by 
abandoning the simultaneous locality with 
respect to both time and space.  
In the previous section, we have seen that 
such a weakening of locality does not 
directly contradict the analyticity of the 
S-matrix.  Also, the argument in the 
first subsection of the present section shows 
that there is in principle a regime where the 
time scale associated with the black hole 
can be much smaller than the 
string scale $\ell_s$ in the srong string-coupling regime. 

The proposed general 
form (\ref{entropy1}), and in particular its lower bound (\ref{entropy2}), 
 suggests that to decode the information, 
it is in general necessary to make the time resolution large by appropriately 
averaging over the time scale, 
in accordance with a viewpoint expressed in Ref.\cite{sussflatads} in the context of Matrix theory. 
The time averaging in turn liberates the 
information stored in the spatial uncertainties and hence 
reduces the value of the entropy. For an observer outside a black-hole horizon, 
decoding all the information stored inside requires 
an observation of infinitely long time. 
 
In connection with holography, we finally remark on the 
connection of the space-time uncertainty relation with 
the so-called UV-IR correspondence,\cite{susswitten} 
which is familiar in the recent literature of 
AdS/CFT correspondence.\cite{maldacena}
In brief, the UV-IR correspondence asserts that the 
UV behavior of the Yang-Mills theory (CFT) on the boundary 
corresponds to the IR behavior of supergravity 
in the bulk, and vice versa.  
On the other hand, the space-time uncertainty relation 
for open strings mediating D-branes leads to a 
similar relation that a small spatial uncertainty 
$\Delta X$ in the bulk corresponds to large uncertainties 
$\Delta T$ along the time-like direction on the brane at the boundary. 
Thus, the space-time uncertainty relation and 
the UV-IR correspondence seem to be equivalent in the sense 
that UV and IR are 
correlated in the bulk and boundary.  However, 
with a little scrutiny, we see that 
there is a small discrepancy in that 
the  UV-IR relation is a statement involving 
classical supergravity and consequently that it only 
requires a macroscopic scale characterized by the 
curvature near the horizon, which is given as 
$R_{ads}\sim (g_sN)^{1/4}\ell_s$. In contrast with this, 
the space-time uncertainty relation only involves the 
string scale $\ell_s$. This puzzle can be resolved as suggested essentially in Ref.\cite{polpeet}
 if we convert the uncertainty along the time-like direction into 
a spatial uncertainty on the brane at the boundary. 
Since, for the brane, open strings 
behave as electric sources, the uncertainty $\Delta T$ with respect to 
time is translated typically into a 
self energy associated with the 
spatial uncertainty $\Delta X_{{\rm brane}}$ within the brane as 
\EQ
\Delta T \sim \Delta X_{{\rm brane}}/\sqrt{g_sN}  ,
\label{conversion2}
\EN
by using the well known fact that the effective Coulomb coupling for 
the superconformal Yang-Mills theory is  $(g_sN)^{1/4}\sim 
(g_{YM}^2N)^{1/4}$. 
This leads to 
$
\Delta X_{{\rm bulk}}\, \Delta X_{{\rm brane}} \sim R_{{\rm ads}}^2
$
which is the relation, involving only the supergravity scale $R_{{\rm ads}}$ actually used in Ref.\cite{susswitten}$^,$\cite{polpeet}, 
for a derivation of the holographic bound for the entropy of D3-branes. 
Note that here we are using the standard AdS coordinate 
used in Ref.\cite{maldacena} instead of that of Ref.\cite{susswitten}. 
The infrared cutoff of order $\Delta X_{bulk}\sim R_{ads}$ 
amounts to an ultraviolet cutoff of order 
$\Delta X_{brane}\sim R_{ads}$ for D-branes at the 
boundary.  
For D3 branes wrapping around a 3-torus of volume $L^3$, 
the degrees of freedom are then 
$
N_{{\rm dof}}\sim N^2L^3/ R_c^3=L^3R_c^5/ G_{10}. 
$

We emphasize that 
the holography and UV-IR correspondence are  of macroscopic nature, involving only 
macroscopic parameters in their 
general expressions.  In fact, the black-hole entropy bound 
and the more general holographic bound have been 
argued (see Ref.\cite{beck} and references therein) 
to follow from the second law of 
thermodynamics, generalized to gravitating systems. 
In contrast to this, the 
space-time uncertainty relation is a general principle 
of a microscopic nature, characterized directly by the 
string scale without any macroscopic variables. 
Hence, in applying the space-time uncertainty relation 
to macroscopic physics, it is in general necessary 
to make appropriate conversions of the scales in various ways,  
depending on different physical situations,  as exemplified typically 
by  (\ref{entropy1}), (\ref{conversion1})  and (\ref{conversion2}). 
The qualitative conformity of the microscopic space-time uncertainty 
relation with holography suggests that the former can be a consistent 
microscopic principle which underlies the required 
macroscopic properties.  As emphasized above, 
the departure of string theory 
from  the framework of 
local field theory seems to be minimal in its nature. But 
the nonlocality of string theory, as being represented by the 
space-time space-time uncertainty principle, appears to 
be sufficient for coping with black-hole complementarity 
and holography.    

Finally, in connection with the problem of macroscopic black holes, 
there remains one big problem. That is the problem of space-time singularities. 
Customarily, we expect that classical 
geometry breaks down around the length scale near the 
string scale $\ell_s$. From the point of view 
of the space-time uncertainty relation, however, 
we have to discriminate the scales with respect to 
time and space. 
If we take the typical example of a Schwarzschild black hole, 
the singularity is a space-like region.  Any object 
after falling inside the horizon encounters  the 
singularity within a finite proper time. If one asks precisely at 
what time it encounters the singularity, the time resolution 
of the clock on the object must be sufficiently small. 
But then the space-time uncertainty relation again 
tells us that the locality with respect to the 
spatial direction is completely lost. 
Thus the classical local-geometric 
formulation which the existence of 
singularity relies upon loses its validity.  Similarly, if 
the singularity is time like, the locality along the time direction 
is completely lost.  It seems thus certain that in string theory 
space-time singularities are resolved.  
However, it is unclear whether this way of resolving the problem of  space-time singularities has any observable significance, characterizing 
string theory.\footnote{
An interesting point is that, in both the cases of the black-hole horizon 
and the space-time singularity,  the 
increase of spatial extendedness of 
strings in the short time limit is coincident with those of 
the spatial distances between the geodesic trajectories exhibited in  the classical 
Schwarzschild metric.  
}

\section{Toward  a 
noncommutative geometric formulation}
 
We have emphasized the role of world-sheet conformal 
symmetry as the origin of the space-time uncertainty relation. 
As has already been alluded to 
at the end of \S 2.2, such a dual relation between 
time and space obviously suggests some mathematical 
formalism which exhibits noncommutativity between operators 
associated to space and time. However, the usual world-sheet 
quantum mechanics of strings does not, at least manifestly, 
exhibit such noncommutativity. 
In a sense, in the ordinary world-sheet formulation,  
use is made of a representation in which the time (center-of-mass time of a string) is diagonalized, 
and the spatial extension $\Delta X$ is measured by the 
Hamiltonian, as is evident in our first intuitive 
derivation of the space-time uncertainty relation. 
Thus the noncommutativity of 
space and time is indeed there in a hidden form. 
 Are there any alternative 
formulations of string quantum mechanics 
which explicitly exhibit 
noncommutativity?  Note that we are not 
asking a further extension of string theory 
with an additional requirement of space-time noncommutativity. 
What is in mind here is a different representation 
of string theory with manifest noncommutativity that is,  
however, equivalent, at the level of the on-shell S-matrix, 
to the  usual formulation, at least perturbatively.  
A different representation may well be more 
suited for an off-shell non-perturbative formulation, hopefully.  
 
The purpose of this section is to suggest a particular possibility 
in this direction.  From the above consideration, 
we should expect the existence of a world-sheet 
picture which is quite different from the ordinary 
one with respect to the choice of gauge.   
Let us consider the so-called Schild action\footnote{
The original action proposed in Ref.\cite{schild} did not 
contain the auxiliary field $e$. However, an equivalent 
condition was imposed by hand. 
}
of the form $(\lambda=4\pi\alpha' , \xi=(\tau, \sigma) )$
\EQ
S_{schild} =-{1\over 2}\int d^2\xi\, e \Bigl\{{1\over e^2}
\Big[-{1 \over 2\lambda^2}(\epsilon^{ab}\partial_a X^{\mu}\partial_b  X^{\nu})^2\Big] + 1
\Bigr\} +\cdots
\label{Schildaction}
\EN
where $e$ is an auxiliary field necessary to maintain 
the reparametrization invariance. 
We  only consider the bosonic part for simplicity.  
The relevance of this action to the space-time 
uncertainty relation has  been discussed 
in a previous work \cite{yo3}  
from a slightly different context. There, it is shown 
how to transform the action into the more familiar Polyakov 
formulation.  Also  the study of this action motivates the 
definition of a particular matrix model, called `microcanonical 
matrix model', as a tentative 
nonperturbative formulation, by introducing a matrix 
representation of the commutation constraint 
(\ref{cconstraint}). This model is quite akin to 
the type IIB matrix model.\cite{ikkt} 

     From the point of view of conformal invariance, the equivalence 
of this action with that of 
the ordinary formulation is exhibited by 
the presence of the same Virasoro condition as the usual one. 
We can easily derive it in the form of
 constraints in the Hamiltonian formalism:
\EQ
{\cal P}^2 + {1\over 4\pi\alpha'}{\acute X}^2 = 0 , \quad 
{\cal P}\cdot {\acute X} =0 .
\label{virasoro}
\EN
In deriving this relation, it is essential to use the condition  
coming from the variation of the auxiliary field $e$,  
\EQ
 {1\over e}\sqrt{-{1\over 2}(\epsilon^{ab}\partial_a X^{\mu}\partial_b  X^{\nu})^2}=\lambda 
\label{conformalconstraint}
\EN
which we proposed to refer to as a `conformal constraint' in Ref.\cite{yo3}.  
Under these circumstances, we can proceed to the ordinary quantization 
with the Virasoro constraint as a first class constraint. 
In this case, there is apparently no place where the noncommutativity 
of space-time coordinates appears.  The 
space-time uncertainty relation is embodied in 
conformal invariance which is typically 
represented by the Virasoro condition.  

Now let us change to another possible representation 
of the Schild action by introducing a new auxiliary 
field $b_{\mu\nu}(\xi) $, which is a space-time antisymmetric tensor of second rank but is also a world-sheet density:
\EQ
S_{b} =-{1\over 2}\int d^2\xi\, e \Bigl\{{1\over e^2}
\Big[{1 \over \lambda^2}\big(\epsilon^{ab}\partial_a X^{\mu}\partial_b  X^{\nu}
b_{\mu\nu} + {1\over 2}b_{\mu\nu}^2\big)\Big] + 1
\Bigr\} .
\label{baction1}
\EN
This can further be rewritten by making the rescaling 
 $b_{\mu\nu}\rightarrow eb_{\mu\nu}$ of the $b$ field:
\EQ
S_{b2} =-\int d^2\xi\, \Bigl\{ {1\over 2\lambda^2}
\epsilon^{ab}\partial_a X^{\mu}\partial_b  X^{\nu}
b_{\mu\nu} + {1\over 2}e\Big({1\over 2\lambda^2}b_{\mu\nu}^2 +1\Big) 
\Bigr\} .
\label{baction2}
\EN
Note that the $b$ field is then a world-sheet scalar. 
Usually, this Lagrangian is not convenient for quantization,  
since it contains only first derivatives with respect to 
the world-sheet (proper) time, leading to second class constraints, 
and there is no kinetic term and no Hamiltonian. 
From the viewpoint of 
noncommutative space-time coordinates, on the other hand, 
the second class 
constraints making identifications between
 some components of momenta and 
coordinates, could be the origin of the noncommutativity. 
If we assume for the moment that the 
external $b$ field is independent of the world-sheet time, 
the Dirac bracket taking account the second class constraint 
is 
\[
\{
X^{\mu}(\sigma_1), X^{\nu}(\sigma_2)
\}_D= \lambda^2 ((\partial_{\sigma}b(\sigma_1)^{-1})^{\mu\nu}
\delta(\sigma_1 -\sigma_2) .
\] 
To see that this conforms to the space-time uncertainty relation, 
it is more appropriate to rewrite it as 
\EQ
\Bigl\{
X^{\mu}(\sigma_1), {1 \over \lambda} \partial_{\sigma}b^{\mu}_{\nu}(\sigma_2)X^{\nu}(\sigma_2)
\Bigr\}_D= \lambda 
\delta(\sigma_1 -\sigma_2) .
\label{diracbracket}
\EN
Since the $b$ field satisfies the constraint equation 
\EQ
{1\over 2\lambda^2}b_{\mu\nu}^2 =-1 ,
\label{bconstraint}
\EN 
assuming 
that the auxiliary field $e$ is first integrated over,  
 we must have nonvanishing 
time-like components $b_{0i}$ of order $\lambda$:
\[
b_{0i}^2 = \lambda^2 + {1\over 2}b_{ij}^2\ge \lambda^2 .
\]
Then (\ref{diracbracket}) is characteristic of  
the noncommutativity between the target time and the space-like 
extension of strings. 

In the general case of a time dependent auxiliary field $b$, 
it is not straightforward to interpret the above action 
within the ordinary framework of canonical quantization, 
since the system is no longer a conserved system, with explicit 
time dependence in the action. 
However, the essence of the noncommutativity  lies 
in the presence of the phase factor itself,  
\[
\exp \big[i\int d^2\xi\, {1\over 2\lambda^2}
\epsilon^{ab}\partial_a X^{\mu}\partial_b  X^{\nu}
b_{\mu\nu}\big] , 
\]
rather than a formal interpretation in terms of operator algebra.  
The path integral in principle contains all the  information of both the 
operator algebra and its representation. Let us assume 
the appearance of this phase factor is an indispensable part of any quantization based on the action (\ref{baction2}). 
Then, we can qualitatively see a characteristic noncommutativity between 
time and space directions directly in this phase factor for the general case. 
To avoid a complication associated with the boundary 
we restrict ourselves to closed strings in the following 
discussion. 

First, in the presence of this phase factor, 
the most dominant configurations for the 
$b$ field for a generic world-sheet configuration 
of the string coordinates are those with the smallest possible 
absolute values allowed under the constraint (\ref{bconstraint}). 
This is because the cancellation 
of the path integral over the world-sheet coordinate 
becomes stronger as the absolute value of $b$ increases. 
So let us first consider the case where the spatial components 
are zero: $b_{ij}=0$, leading $b_{0i}^2=\lambda^2$. 
The effect of the spatial components $b_{ij}$, 
corresponding to the noncommutativity 
among spatial coordinates, will be  
briefly described later. 
Under this approximation, dependence on the world-sheet 
coordinate in the $b$ field satisfying the constraint
 can be expressed as a local $O(D-1)$ rotation 
belonging to a coset $O(D-1)/O(D-2)$:
\EQ
b_{0i}(\tau, \sigma)=\lambda S_{ri}(\tau, \sigma) .
\label{localrotation}
\EN
Here we represent the coset element by the matrix elements 
$S_{ri}$, with $r$ being the radial direction for definiteness.  

 Let us now choose the time-like gauge   
\[
\partial _{\sigma}X^0=0
\]
and treat the target time 
as a globally  defined dynamical variable on the 
world-sheet as a function of the world-sheet time parameter 
$\tau$. 
Then the phase factor reduces to  
\[
\exp \big[i\int d\tau  {1\over \lambda^2}
\dot{X}^0 \int d\sigma b_{0i}(\xi) \partial_b  X^i
\big] .
\]
We can interpret this phase factor as arising from the 
product of the short-time (with respect to world-sheet time)  
matrix elements 
\EQA
\Big\langle X^0&&(\tau + {1\over 2}\epsilon)\Big|
X^0(\tau -{1\over 2}\epsilon)\Big\rangle  \nonumber \\
&&=\int [d\vec{X}dS(\tau,\sigma)]
 \Big\langle  X^0(\tau + {1\over 2}\epsilon)\Big| \vec{X}, 
\partial_{\sigma}S\Big\rangle 
\Big\langle  \vec{X}, 
\partial_{\sigma}S\Big|
X^0(\tau -{1\over 2}\epsilon)\Big\rangle 
\EQN
where the intermediate state to be integrated over  
is inserted at the mid-point and the matrix elements are 
\EQ
 \Big\langle  X^0(\tau + {1\over 2}\epsilon)\Big| \vec{X}, 
\partial_{\sigma}S\Big\rangle
=\exp \big[i{1\over \lambda^2}
X^0\big(\tau+{1\over 2}\epsilon\big) \int d\sigma b_{0i}(\xi) \partial_b  X^i
\label{phase1}
\big] 
\EN
\EQ
\Big\langle  \vec{X}, 
\partial_{\sigma}S\Big|
X^0(\tau -{1\over 2}\epsilon)\Big\rangle =
\exp \big[-i{1\over \lambda^2}
X^0(\tau-{1\over 2}\epsilon) \int d\sigma b_{0i}(\xi) \partial_b  X^i
\big] .
\label{phase2}
\EN
In a more familiar operator form, this would correspond to the commutator 
\[
\Big[X^0, \int d\sigma \, S_{ri} \partial_{\sigma}  X^i \Big]=i\lambda
\label{timespacecomm}
\]
at each instant of the world-sheet time. But the phase factors,  as 
exhibited in (\ref{phase1}) and (\ref{phase2}), 
 lead directly to an uncertainty relation 
of the form
\EQ
|\Delta X^0||\Delta \vec{X} | \gtsim \lambda 
\EN
\EQ
 |\Delta \vec{X}| = \sqrt{\Big\langle \Big(\Delta \int d\sigma S_{ri} \partial_{\sigma}  X^i(
\sigma) \Big)^2\Big\rangle }
\label{spatialuncertainty}
\EN
with respect to the orders of magnitude of uncertainties in the path integral, by the same mechanism as the 
ordinary Fourier transformation.  We note that (\ref{spatialuncertainty}) is 
invariant under reparametrization with respect to $\sigma$. 
Furthermore,  the latter is 
acceptable as a measure for the spatial uncertainty, 
since it locally measures the length along the 
tangent of the profile of closed strings at a fixed world-sheet 
time, including the possibility of multiple 
winding, provided it does not vanish. 
In particular, when $\partial_{\sigma} X^i(\sigma)$ and $S_{ri}(\sigma)$ 
as two vectors in the target space 
are parallel to each other along the string, 
it precisely agrees with the proper length measured along the 
string. For general random configurations of 
the orientation of these vectors, 
(\ref{spatialuncertainty}) is a possible general definition 
of the length of a string in a coarse-grained form.  

The effect of spatial 
components $b_{ij}$ can be taken 
into account if we generalize the local rotation to the local 
Lorentz group $O(D-1, 1)$ in (\ref{localrotation}). 
This is due to the fact that we can restrict the 
components of the auxiliary field $b$ to those which 
have nonvanishing product $b_{\mu\nu}\epsilon^{ab}\partial_a X^{\mu}\partial_b  X^{\nu}$. Since the antisymmetric tensor 
$\sigma_{\mu\nu}=\epsilon^{ab}\partial_a X^{\mu}\partial_b  X^{\nu}/2
$ can  be  locally transformed to one corresponding to a 
time-like two-dimensional plane,\footnote{
In terms of invariants, this corresponds to the following 
property. If $\sigma^2_{\mu\nu}=\Sigma_{\mu\nu}$, then  
$\Sigma^2_{\mu\nu}=\Tr(\Sigma)\Sigma_{\mu\nu}/2$. 
Thus there is only one independent Lorentz invariant. 
} we can assume 
a parametrization, say $b_{\mu\nu}=\lambda S_{0\mu}S_{r \nu}$,  
using the rotation matrix of $O(D-1,1)$. 
This leads to a correction to the definition of the 
spatial uncertainty as 
\[
 |\Delta \vec{X}| = \sqrt{\Big\langle \Big(\Delta \int d\sigma 
(S_{00}S_{ri}-S_{0i}S_{r0}) \partial_{\sigma}  X^i(
\sigma) \Big)^2\Big\rangle } .
\]
Also, there arises an induced noncommutativity among the 
spatial components, corresponding to the 
phase factor
\[
 \exp \big[i{1\over \lambda^2}
 \int d\tau d\sigma  \, \dot{X}^iX'^j(S_{0i}S_{rj}-S_{0j}S_{ri})\big] .
\]
This should be interpreted as  residual noncommutativity,  which is necessary to preserve 
Lorentz invariance in the presence of the primary 
noncommutativity between time and space. 

Although a more rigorous formulation is desirable, 
our discussion seems to already suggest the quite remarkable 
possibility that  space-time noncommutativity alone 
 governs the essential features of the dynamics. This would not be 
so surprising if we recall that the space-time 
uncertainty relation can be regarded as a 
reinterpretation of the time-energy uncertainty relation. As such, 
 its proper formulation would necessarily amount to 
formulating the Hamiltonian appropriately, 
as should have been clear from our foregoing 
discussions.

Of course, this particular formalism does not seem convenient  
for performing concrete computations of 
string amplitudes, at least with the technical tools 
presently available to us.  
Also, our discussion, being based on 
the world-sheet picture, is still perturbative in its nature. 
As we have stressed, the 
space-time uncertainty relation should be valid nonperturbatively, 
and hence must be ultimately reformulated  without relying 
on the world-sheet picture on the basis of some 
framework which is second-quantized from the 
outset. The connection with matrix models 
discussed in a previous work \cite{yo3} is certainly suggestive 
of a nonperturbative formulation, but 
 unfortunately it seems to be yet lacking  some key ingredients for 
a definitive formulation.  
We hope, however,  that the above argument gives some impetus for 
further investigation aimed at constructing truly  nonperturbative and 
calculable formulations in the future. 
For example, from the viewpoint of an analogy between classical 
phase space and space-time that we mentioned in discussing 
the generalized conformal transformation, 
 study of the most general transformations  
which preserve the form $i\int d^2\xi\, {1\over 2\lambda^2}
\epsilon^{ab}\partial_a X^{\mu}\partial_b  X^{\nu}
b_{\mu\nu}$ might be a direction to be pursued. 

       In connection with this, it might  be 
possible to reinterpret directly the action (\ref{baction2}) as a 
generalized deformation 
quantization of space-time geometry itself. 
This expectation also suggests 
a formulation from the viewpoint of M-theory 
by interpreting the world sheet of strings as a 
section of a membrane, and using a sort of formalism  
 related to the Nambu bracket.\cite{nambu}  We also mention 
that to make the comparison with local field theory, 
the approach suggested in Ref.\cite{kato} might be 
of some relevance in the case of open strings.  
We have left all these possibilities as  
challenging and promising problems for the future.    

The reader may have noticed the similarity 
regarding
 the appearance of noncommutativity 
in the present discussion with that in the 
recent discussions of noncommutative Yang-Mills theory 
based on D-branes. An obvious difference is 
that our $b$ field is a world-sheet field 
which always exist even without the presence of the 
external space-time  $B$ field. 
Note that we  obtained noncommutativity in the sense of 
the target space-time from the world sheet $b$ field in the bulk of the 
string world sheet.  But this noncommutativity 
is simply another representation 
of the space-time uncertainty relation 
already exhibited in the usual 
formulation with manifest conformal symmetry. 
Also, in our case, the dominant components of the $b$ field 
are the time-like components $b_{0i}$, in contrast to the 
space-like components of the $B$ field 
in Refs.\cite{douglas} and \cite{seiwitten}.  
If we had treated D-branes using 
the above formulation based on the 
Schild action 
for open strings attached to D-branes by adding the 
constant space-time $B_{ij}$ field,  
we would have obtained the noncommutativiy 
between time and space directions as above along the 
D-brane world volume, 
in addition to the noncommutativity among spatial directions along 
D-branes in association with $B_{ij}$. 

Our approach to noncommutativity is  also quite different from that 
of Ref.\cite{ishibashi} in type-IIB matrix models. 
However, since the Schild action is intimately 
connected to the type-IIB matrix model, 
it would be very interesting to seek some possible 
relation with it. 
 
We emphasize again 
that the noncommutativity discussed in the 
present section 
between time and space 
 is a property which is intrinsic 
to the dynamics 
of fundamental strings, and it has  nothing to do with 
the presence or absence  of the external $B$ field. 
Of course, the space-time $B$ field is automatically contained 
as a state of closed strings in any valid formulation 
of (orientable) string theory.  In quantum theory, 
we have to take into account  its vacuum 
fluctuations.  In this broad sense, these two 
different origins of space-time noncommutativity
 might be united in some nonperturbative framework, 
by identifying the fluctuations of the space-time $B$ field and 
the world sheet $b$ field self-consistently.

\section{Further remarks} 
In this final section, we discuss some miscellaneous points 
which have not been treated in the preceding sections and 
may  become the source of confusion. We also 
comment on some future possibilities.  

\vspace{0.2cm}
\noindent
{\it Frame dependence, $(p, q)$ strings, and  S-duality}

\vspace{0.1cm}
Since the space-time uncertainty relation 
is a statement which contains a dimensionful parameter 
$\ell_s$, we have to specify the frame for the metric in the sense of the 
Weyl transformation, with respect to 
which the string length parameter is defined. 
In the foregoing discussions, we always tacitly 
assumed that the string length $\ell_s$ is the proportional constant  
in front of the world-sheet string action, say 
$(1/\ell_s^2)\int d^2 \xi g_{\mu\nu}\partial_z X^{\mu}
\partial_{\bar{z}}X^{\nu} +\cdots $, using the standard 
conformal gauge. 
Therefore the frame of the space-time 
metric $g_{\mu\nu}(X)$ which should be used for 
the space-time uncertainty relation is the 
so-called string frame metric. 
This is important when we consider the S-duality transformation, 
under which the string-frame  metric is not invariant. 

Suppose we start with the fundamental string [$(1,0)$ 
string] in type IIB theory and make a S-duality transformation 
which send  $(1,0)$ strings to  $(p, q)$ strings. 
In the original (1,0) picture, the 
other $(p, q)$ strings are soliton excitations. 
Therefore, their interaction and motion are governed 
by the fundamental strings. In this sense, the space-time 
uncertainty relation must be satisfied using the original 
string frame metric at least in the weak coupling regime,  
where the tension of the (1,0) string is 
smaller than the $(p, q)$ strings, provided we correctly identify 
the uncertainties. 
Note that the same can also be said for other higher dimensional D-branes.\footnote{
For a discussion of some uncertainty relations  
along the D-brane world volume, see Ref.\cite{chuho}.  
} 
As long as we consider them in the weak string coupling regime 
with respect to the original fundamental string,  all  of 
the dynamics are basically expressible in terms of the fundamental 
strings. Although we now know that string theory is full 
of objects of various dimensions, they cannot be 
treated  in a completely democratic way from the 
point of view of their real dynamics.   

However, if we wish to use the 
picture in which the $(p, q)$ string is now treated as the 
fundamental string in the regime where the transformed 
string coupling $g_s^{(p,q)}=\exp \phi_{(p,q)}$ is weak,  
and hence the original string coupling is in general in 
a strong-coupling regime, 
we have to use the world-sheet action of the 
$(p,q)$ string to describe the dynamics. 
Then it is essential to 
shift our space-time Weyl frame correspondingly. Namely, 
the space-time string metric must also be transformed by the 
same S-duality transformation. This precisely cancels 
the difference of tensions for $(1,0)$ and $(p, q)$. 
This is, of course,  as it should be as long as the 
S-duality transformation is a {\it symmetry} of the 
type IIB superstring theory. The space-time uncertainty 
relation is therefore invariant under the S-duality transformation. 
Thus, at least in S-duality symmetric theories, 
the space-time uncertainty relation must be valid 
for arbitrary string coupling, provided the appropriate 
change of the Weyl frame is made according to the 
transformation law of S-duality and the uncertainties are 
redefined correspondingly.  

In formulas, this can be expressed as follows.  The world-sheet bosonic action 
for the $(p,q)$ string is, using the 
ordinary string metric of the target space-time as 
the fundamental $(1,0)$ string, 
\[
T_{(p,q)}\int d^2 \xi \, g_{\mu\nu}(X) \partial_z X^{\mu}
\partial_{\bar{z}}X^{\nu} ,
\]
where the tension of the $(p,q)$ string in the original string frame units is 
given by \cite{schwarz} 
\EQ
T_{(p,q)}=\bar{\triangle}^{1/2}_{(p,q)}{1\over \ell_s^2} ,
\EN
and 
\EQ
\bar{\triangle}_{(p,q)}=|p-q\rho|^2 = \exp(\phi_{(p,q)}-\phi_{(1,0)})  ,
\EN
with $\rho = {\chi\over 2\pi} +ie^{-\phi}$. 
On the other hand, the space-time string metrics 
are related by 
\[
g_{\mu\nu}^{(p,q)}(X) \exp(-\phi_{(p,q)}/2)=
g_{\mu\nu}^{(1,0)}(X) \exp(-\phi_{(1,0)}/2) ,
\]
with $g_{\mu\nu}=g_{\mu\nu}^{(1,0)}$,  
corresponding to the S-duality invariance of the 
Einstein frame metric. 
Combining these relations, we confirm that 
the world-sheet action of the $(p,q)$ string is 
equal to
\[
{1\over \ell_s^2}\int d^2 \xi \, g^{(p,q)}_{\mu\nu}(X) \partial_z X^{\mu}
\partial_{\bar{z}}X^{\nu} . 
\]
Thus we have a space-time uncertainty relation with the 
same string length $\ell_s$ as that before making the 
transformation. 

\vspace{0.2cm}
\noindent
{\it Curved or compactified space-time and a remark on T-duality}

\vspace{0.1cm}
Another point related to that discussed above is that 
the space-time uncertainty relation must be valid qualitatively 
in general curved space-times allowed as backgrounds 
of string theory, 
as long as the world-sheet conformal invariance is 
not violated. 
In this case too, it is essential to use the 
string frame metric to measure the invariant (or proper) length 
appropriately with respect to time and space directions.\footnote{
For example, the discrepancy claimed in Ref.\cite{bakrey} 
can easily be corrected by using the proper length 
appropriately. 
} 

A somewhat related, but different point involves the interpretation of T-duality  from the viewpoint of the 
space-time uncertainty relation. 
T-duality asserts that under the compactification of a spatial 
direction along a circle, the theory 
with a radius $R$  is equivalent to that with 
$\ell_s^2/R$.  This is due to the mapping  
$n\rightarrow m, R\rightarrow \ell_s^2/R$ between 
the momentum modes whose mass spectrum 
is $n/R$ and the winding modes whose spectrum is 
$m R/\ell_s^2$.  From the viewpoint of the space-time 
uncertainty relation, the uncertainty with respect to the 
former, referring only to the center-of-mass 
momentum,  must be translated into an uncertainty with respect to 
energy by
\[
\Delta T_1 \sim R_1/\Delta n_1 
\] 
which implies the lower bound for the spatial uncertainty 
$\Delta X_1\sim \ell_s^2\Delta n_1/R_1$.
Here we have used  the label $1$ to denote the uncertainty relation 
in theory $1$. 
Suppose that theory 1 is mapped into theory 2,  which is compactified 
with a radius $R_2$, 
by identifying the spatial uncertainty 
$\Delta X_1\rightarrow \Delta X_2=R_2\ell_s\Delta m_2$ 
originating from the uncertainty 
with respect to the winding number, giving 
$\Delta T_2\sim \ell_s^2/R_2\Delta m_2$. 
Thus the uncertainty relations of the two theories are
 related to 
each other by making the mapping 
\[
n_1\rightarrow m_2, \quad m_1 \rightarrow n_2, \quad 
R_1 \rightarrow \ell_s^2/R_2  .
\]
This is precisely the mapping of the T-duality transformation. 
Thus T-duality is consistent with the space-time 
uncertainty relation, as it should be. In connection with 
this, it must be kept in mind that for the uncertainty with respect to
spatial directions, we have to take into account windings. 
For example, the definition of the spatial uncertainty suggested 
from the Schild action, as 
discussed in the last section,  indeed naturally contains the 
winding effect.  
Another remark is that, in our interpretation, T-duality is a statement regarding duality between small and large distances 
in time and space directions, rather than regarding the 
existence of a minimal distance as is often expressed in the 
literature.  

\vspace{0.2cm}
\noindent
{\it The role of supersymmetry ?}

\vspace{0.1cm}
In our discussions, the space-time supersymmetry has not 
played a fundamental role.  The reason is that the supersymmetry 
is not directly responsible for the short distance structure of 
string theory.  It rather plays a central role 
in ensuring the theory be well defined, at least perturbatively, 
in the long distance regime. However, 
the space-time uncertainty relation essentially demands 
dual roles between ultraviolet and infrared regimes 
by interchanging the temporal  and spatial directions. 
In this sense,  the space-time supersymmetry 
actually  plays an important subsidiary role in order to make 
the theories well-defined in both ultraviolet and infrared 
regimes.  Such an instance was already explained 
for the case of D-particle dynamics. 

In connection with this, a question arises whether we have to 
impose, in future 
nonperturbative formulations of string theory,  supersymmetry as an additional assumption 
which is not automatically guaranteed from 
fundamental principles alone.   Although we do not know the answer, 
recent developments \cite{sen} on unstable D-brane systems 
 indicate that  the mere appearance of tachyons  
should no more be regarded as the criterion of 
unacceptable theories.  This only signifies that 
the perturbative vacuum we have chosen to start with 
is wrong. Indeed, it was recently shown  by the  
present author \cite{yosusy} 
that the 10-dimensional (orientable) open string theory 
with both bosons and fermions,
either its Neveu-Schwarz-Ramond or Green-Schwarz 
formulation,   has a hidden $N=2$  space-time supersymmetry automatically without making  the standard GSO projection.  
It is an important question whether a similar 
interpretation is possible for closed string theories as well.

\vspace{0.2cm}
\noindent
{\it M-theory interpretation of the space-time uncertainty relation ?}

\vspace{0.1cm}
Let us next reconsider the relevance of the 
space-time uncertainty principle to the M-theory 
conjecture. In \S 4, we  derived the 
M-theory scale from two different points of view, 
namely those of microscopic black holes in 10-dimensional 
space-times and of D-particle dynamics. In particular, the 
former argument shows that the appearance of the 
M-theory scale can be a quite general phenomenon, 
not necessarily associated with D-branes.   

One of the basic elements of the M-theory 
conjecture is that in 11 dimensions the role 
of fundamental strings is replaced by that of  membranes, 
which are wrapped around the compactified 
circle of radius $R_{11}=g_s\ell_s$. 
From this point of view, it seems natural \cite{liyo2}$^,$\cite{minic} to 
further reinterpret the space-time uncertainty 
relation as 
\EQ
\Delta T \Delta X \gtsim \ell_s^2 \sim \ell_M^3/R_{11} 
\, \, \rightarrow \Delta T \Delta \vec{X} \Delta X_{11}
\gtsim \ell_M^3 \sim G_{11}
\label{mstu}
\EN
by setting $\Delta X_{11}\sim R_{11}$ as the 
uncertainty along the 11-th direction and taking $\Delta X \rightarrow 
\Delta \vec{X}$ which is identified as
 the spatial uncertainty in the 
9 dimensional transverse directions.  
This is in accord with the membrane action which has two space-like directions along the 
world volume of membrane. In Ref.\cite{liyo2}, 
we have discussed the affinity of this relation 
with AdS/CFT correspondence in 11 dimensions. 
This also motivated the study of the Nambu bracket in Ref.\cite{almy}. 
The original stringy space-time uncertainty 
relation would then be an approximation of this relation in 
the limit of small compactification radius. 
Once we move to this viewpoint, the 
fundamental scale is 
now $\ell_M=\ell_{11}\sim g_s^{1/3}\ell_s$.  Of course, any genuinely 
11 dimensional effects only appear 
for large compactification radius,  
$R_{11}\gg \ell_M $. In this regime, all the 
characteristic scales of the theory are 
governed by the order $\ell_M$. 
The appearance of different scales for time 
and space scales in 10 dimensions controlled 
by the string coupling is obviously an 
effect of the small compactification scale $R_{11}\ll \ell_M$. 

For example, we can apply the 
same argument for microscopic 
black holes to derive the criterion determining 
where truly M-theory effects take place. 
The black hole uncertainty relation 
places a restriction in 11 dimensions as 
\EQ
\Delta T (\Delta X)^{8} \gtsim \ell_M^9 .
\label{mbhstu}
\EN
Comparing with the M-theory uncertainty relation 
(\ref{mstu}), we find that the critical point 
is of the same order, 
\[
\Delta T_c \propto \Delta X_c \propto \ell_M , 
\]
if we treat all the spatial directions equivalently. This 
 is more or less evident from the outset since there can be no other scales 
than $\ell_M$  
unless one puts them in by hands.  
Therefore in this case, the dimensionless proportionality 
coefficients are very important in order to 
ascertain various characteristic scales. 
In this sense, in M-theory, understanding 
 the real nonperturbative mechanisms for 
generating the low-energy scales becomes 
completely nonperturbative at a much higher level 
than in 10 dimensional string theory. 

We also note that it is straightforward to 
extend the Schild action approach introduced in \S 5 
to a  noncommutative geometric formulation for the 
quantization of a membrane.  
In this case, the role of the world-sheet auxiliary field 
$b_{\mu\nu}$ is played by a world-volume 
3 rank tensor field $c_{\alpha\beta\gamma}(\xi)$. 
We can easily derive an analog  
of the stringy uncertainty 
(\ref{spatialuncertainty}) for membranes.  

Quite recently, it has been 
argued \cite{li} that the relation (\ref{mstu}) is 
compatible with the so-called 
`stringy exclusion principle' \cite{mastro} on  
AdS space-times, by reinterpreting an 
observation made in Ref.\cite{macsuss}.  Also, an 
approach proposed in Ref.\cite{antalsanje} to the 
stringy exclusion principle 
suggests a connection with the quantum 
group interpretation, 
another possible manifestation of noncommutativity,
 of these phenomena. 

\vspace{0.2cm}
\noindent
{\it A fundamental question}

\vspace{0.1cm}
In the beginning of this paper, we repeatedly stressed 
 the importance of reinterpreting 
the role of world-sheet conformal symmetry in terms of 
some new language, which is not in principle 
dependent upon perturbation theory, as a motivation 
of our proposal of the space-time uncertainty principle. 
There, however, remains still 
one of the most mysterious questions in string theory.  Why does string theory 
contain gravity?\footnote{
For a recent general review on this question, see Ref.\cite{beijingyo}. 
}  
 Of course, we have checked the 
consistency of the space-time uncertainty relation with the 
presence of 
gravity from various viewpoints.  In spite of these 
many  checks,  
it is still unclear what ensures the 
appearance of general relativity in the long distance regime. 
The main reason for this deficiency is that 
we have not gained an appropriate understanding of the symmetries 
associated with the space-time uncertainty 
principle in terms of the target space-time. 
The generalized conformal symmetry we mentioned 
in \S 4 might contain some ideas which might form a 
germ for investigation in such directions. 

\vspace{0.2cm}

Although many questions still remain,  
summarizing all that  we have discussed in the 
present paper, it seems not unreasonable to assert 
that the space-time uncertainty principle may 
be one of the possible general underlying principles
 governing the 
main qualitative features of string/M theory.  
Of course, the scope of qualitative principles,  
such as our space-time uncertainty principle,  
is much too limited to make any concrete 
predictions without having definite mathematical 
formulations. In this paper, we have tried to clarify  
its meaning and implications as far as 
we can at the present stage 
of development. 
It would be extremely interesting to arrange the 
various aspects discussed here
 into a unified mathematical scheme.    
 
\section*{Acknowledgements}
The present paper was essentially 
completed during the author's visit 
to Brown University in March, 2000. He would like 
to thank the Department of Physics there 
for the hospitality and also discussions with 
D. Lowe, A. Jevicki and S. Ramgoolam during preparation of  
the manuscript.  
The present work  is supported in part 
by a Grant-in-Aid for Scientific  Research (No. 09640337) 
and a Grant-in-Aid for International Scientific Research 
(Joint Research, No. 10044061) from the Ministry of  Education, Science and Culture. 



\begin{thebibliography}{99}
\bibitem{over} For a convenient review of string theory including some 
recent development, see,  e.g.,  
J. Polchinski, {\it String Theory} Vol I \& II, Cambridge Univ. Press. , 1998. 
\bibitem{yo0}T. Yoneya, Lett. Nuovo. Cim.  8(1973)951;Prog. Theor. Phys. 51(1974) 1907; 56(1976)1310.
\bibitem{ss} J. Scherk and J. Schwarz, Nucl. Phys. B81(1974)118;
Phys. Lett. 57B(1975)463.
\bibitem{kk} M. Kaku and K. Kikkawa, Phys. Rev. D10 (1974) 1110, 1823.
\bibitem{wittensft} E. Witten, Nucl. Phys. B268 (1986) 79; B276 (1986) 291.
\bibitem{hikko} H. Hata, K. Itoh,  T. Kugo, H. Kunitomo and 
K. Ogawa, Phys. Rev. D34 (1986) 2360. 
\bibitem{nw} A. Neveu and P. C. West, Phys. Lett. 168B (1986) 192. 
\bibitem{fs} D. Friedan and S. Shenker, Nucl. Phys. B281 (1987) 509.  
\bibitem{banksetal} T. Banks and  E. Martinec, Nucl. Phys. B294 (1987) 733. 
\bibitem{nakanishi} N. Nakanishi, Prog. Theor. Phys. 45(1971) 436. 
\bibitem{yonakadec} T. Yoneya, Prog. Theor. Phys. 
48(1972) 2044.  
\bibitem{giddmar} B. Giddings and E. Martinec, 
Nucl. Phys. B278 (1986) 91. 
\bibitem{oldmatrix} For a review, see {\it e.g.} 
J. Polchinski,  Les Houshes lecture, 
hep-th/9411028 (1994). 
\bibitem{pol} J. Polchinski, 
Phys. Rev. Lett. 75(1995) 4724.
\bibitem{wittenym} E. Witten, Nucl. Phys. B460(1995) 330. 
\bibitem{bfss} T. Banks, W. Fischler, S. H. Shenker and L. Susskind, Phys. Rev. D55 (1997) 5112. 
\bibitem{ikkt} N. Ishibashi, H. Kawai, Y. Kitazawa and A. Tsuchiya,Nucl. Phys B498(1997)467.
\bibitem{yo} T. Yoneya, {\it Duality and Indeterminacy Principle in String  
Theory}
in ``Wandering in the Fields", eds. K. Kawarabayashi and
A. Ukawa (World Scientific, 1987), p. 419: see also
{\it String Theory and Quantum
Gravity} in ``Quantum String Theory", eds. N. Kawamoto
and T. Kugo (Springer, 1988), p. 23.
\bibitem{yo1} T. Yoneya, Mod. Phys. Lett.  A4, 1587(1989).
\bibitem{minimal} See, {\it e.g.}, D. Gross, Proc. XXIV Int. 
Conf. High Energy Physics, Munich, Eds. R. Lotthaus and J. K\"{u}hn, Springer, Verlag (1989). 
\bibitem{shenker} S. Shenker, hep-th/9509132.
\bibitem{grossmende} D. Gross and P. Mende, 
Nucl. Phys. Nucl. Phys. B303 (1988) 407. 
\bibitem{moduncer} For the present author, it is not clear 
what is the original reference for this particular proposal. 
As an example of works in which the relation has been 
stressed, see, {\it e. g.}, R. Guida, K. Konishi and P. Provero, 
Mod. Phys. Lett. A6 (1991) 1487 and earlier references therein. 
\bibitem{wittenm} E. Witten, Nucl. Phys. B443 (1995) 85. 
\bibitem{yo3} T. Yoneya, Prog. Theor. Phys. 97(1997) 949.
\bibitem{dopl} S. Doplicher, K. Fredenhagen and J. E. Roberts, 
Commun. Math. Phys. {\bf 172} (1995) 187; Phys. Lett. {\bf B331} (1994) 39.
\bibitem{alf} See. e.g., L. V.  Alfors,  {\it Conformal Invariants : 
Topics in Geometric Function Theory} (McGraw-Hill, New York,  1973), Chapter 5.
\bibitem{oogurimende} P. Mende and H. Ooguri, 
Nucl. Phys.  B303(1988), 407.
\bibitem{lowe} D. Lowe, Nucl. Phys. B456 (1995) 257.
\bibitem{soldate} M. Soldate, Phys. Lett. 186B (1987) 321.  
\bibitem{amati} D. Amati, M. Ciafaloni and G. Veneziano, Phys.  
Lett. {\bf 197B}(1987), 81. \\
See also an interesting paper by Veneziano prior to this work, G. Veneziano, Europhys. Lett. 2 (1986) 199. 
\bibitem{mueller} A. H. Mueller, Nucl. Phys. B118 (1977) 253.\\ 
See also a related discussion in \cite{suss2} 
based on parton picture. 
\bibitem{celmar} F. Celcius and A. Martin, Phys. Lett. 
8 (1964) 80. 
\bibitem{martinec} E. Martinec, Class. Quan. Grav. 10 (1993) L187. 
\bibitem{polsusslowe} D. Lowe, J. Polchinski, L. Susskind, 
L. Thorlacius and J. Uglam, Phys. Rev. D52 (1995) 6997. 
\bibitem{bhuncertain} For a recent related discussion and 
references for earlier works, see N. Sasakura, 
hep-th/0001161 and Prog. Theor. Phys. 102 (1999) 169. 
\bibitem{thooft2} G. 't Hooft, gr-qc/9608037. 
\bibitem{liyo} M. Li and T. Yoneya, Phys. Rev. Lett. {\bf 78}
 (1997) 1219.
\bibitem{liyo2} M. Li and T. Yoneya, Chaos, Solitons and Fractals
  10(1999) 423--443;  hep-th/9806240. 
\bibitem{dkps} M. Douglas, D. Kabat, P. Pouliot and S. Shenker,
Nucl. Phys. B485 (1997) 85.
\bibitem{bbpt} K. Becker, M. Becker, J. Polchinski and A. Tseytlin, 
Phys. Rev. D56 (1997) 3174. 
\bibitem{okyo} Y. Okawa and T. Yoneya, Nucl. 
Phys. B538 (1998) 67 ;Nucl. Phys. B541 (1999) 163. 
\bibitem{kazamamura} Y. Kazama and T. Muramatsu, hep-th/0003161. 
\bibitem{jeyo} A.  Jevicki and T. Yoneya, Nucl. Phys. B535 (1998) 335.
\bibitem{jekayo} A.  Jevicki, Y. Kazama and T. Yoneya, Phys. Rev. Lett. 81 (1998) 5072 ; Phys. Rev. D59(1999) 066001.
\bibitem{seyo} Y. Sekino and T. Yoneya, Nucl. Phys., in press; 
hep-th/hep-th/9907029. 
\bibitem{yo4} T. Yoneya, Class. Quan. Gravity 17 (2000) 1307.  
\bibitem{sussbanks} T. Banks and L. Susskind, hep-th/9511491. 
\bibitem{seibergsen} N. Seiberg, Phys. Rev. Lett. 79 (1997) 3577.\\
A. Sen, hep-th/9709200.  
\bibitem{sussetal} For a review, see, {\it e.g.} L. Susskind and 
J. Uglam, hep-th/9511257. 
\bibitem{suss2} L. Susskind, Phys. Rev. D49 (1994) 6606. 
\bibitem{horopol} G. Horowitz and J. Polchinski, Phys. Rev. D55(1977) 6189.
\bibitem{thooft} G. 't Hooft, gr-qc/9310026. 
\bibitem{suss3} L. Susskind, J. Math. Phys.  36 (1995)6377.
\bibitem{sussflatads} L. Susskind, hep-th/9901079.
\bibitem{polpeet} A. W. Peet and J. Polchinski, hep-th/9809022.
\bibitem{susswitten} L. Susskind and E. Witten, hep-th/9805114.
\bibitem{maldacena} J. Maldacena, Adv. Theor. Math. Phys. 2 (1998) 231.
\bibitem{beck}  J. Beckenstein, hep-th/0003058. 
\bibitem{schild}  A. Schild, Phys. Rev. {\bf D16} (1977) 1722.
\bibitem{nambu} Y. Nambu, Phys. Rev. Phys. Rev. D7 (1973) 2405.
\bibitem{kato} M. Kato, Phys. Lett. B245 (1990) 43. 
\bibitem{douglas} A. Conne, M. R. Douglas and A. Schwarz, 
JHEP 9802:003 (1998); hep-th/9711162.  
\bibitem{seiwitten}  N. Seiberg and E. Witten, hep-th/9908142. 
\bibitem{ishibashi} H. Aoki, N. Ishibashi, S. Iso, H. Kawai, 
Y. Kitazawa and T. Tada, hep-th/9908141.
\bibitem{chuho} C. S. Chu, P. M. Ho and Y. C. Kao, 
Phys. Rev. D60 (1999) 126003. 
\bibitem{schwarz} J. H. Schwarz, Phys. Lett. B360 (1995) 13. 
\bibitem{bakrey} D. Bak and S. J. Rey, hep-th/9902101. 
\bibitem{sen} A. Sen, JHEP 08 (1998) 010; JHEP 08 (1998) 012.
\bibitem{yosusy} T. Yoneya, Nucl. Phys.,  in press ; hep-th/9912255. 
\bibitem{almy} H. Awata, M. Li, D. Minic, and T. Yoneya, 
hep-th/9906248.  
\bibitem{minic} D. Minic, Phys. Lett. B442 (1998) 102. 
\bibitem{li} M. Li, hep-th/0003173. 
\bibitem{mastro} J. Maldacena and A. Strominger, JHEP 9812 (1998) 005. 
\bibitem{macsuss}  J. McGreevy, L. Susskind and N. Toumbas, hep-th/0003075. 
\bibitem{antalsanje} A. Jevicki and S. Ramgoolam, 
JHEP 9904 (1999) 032. 
\bibitem{beijingyo} T. Yoneya, hep-th/0004075, Proceedings of 
`{\it Frontiers of Theoretical Physics}', ITP, Beijing,  November, 1999. 
\end{thebibliography}
\end{document}